\documentclass[twocolumn,times,twocolappendix]{aastex631}

\newcommand\ci{[C\,{\sc i}] }
\newcommand\cii{[C\,{\sc ii}] }

\newcommand\sersic{S\'{e}rsic }

\usepackage{comment}

\shorttitle{Multiple origins of dusty starbursts in luminous SMGs}
\shortauthors{Ikeda et al.}
\graphicspath{{./}{figures/}}
\usepackage{multirow}
\begin{document}

\title{\large{Formation of Substructure in Luminous Submillimeter Galaxies (FOSSILS): \\ Evidence of Multiple Pathways to Trigger Starbursts in Luminous Submillimeter Galaxies}}

%% you desire. In v6.31 you can include a footnote in the title.

%%
%% Note that \altaffilmark and \altaffiltext have been removed and thus 
%% can not be used to document secondary affiliations. If they are used latex
%% will issue a specific error message and quit. Please use multiple 
%% \affiliation calls for to document more than one affiliation.
%%
%% The new \altaffiliation can be used to indicate some secondary information
%% such as fellowships. This command produces a non-numeric footnote that is
%% set away from the numeric \affiliation footnotes.  NOTE that if an
%% \altaffiliation command is used it must come BEFORE the \affiliation call,
%% right after the \author command, in order to place the footnotes in
%% the proper location.
%%
%% Use \email to set provide email addresses. Each \email will appear on its

%\correspondingauthor{August Muench}
%\email{greg.schwarz@aas.org, gus.muench@aas.org}

\author[0000-0002-2634-9169]{Ryota Ikeda}
\affiliation{Department of Astronomy, School of Science, SOKENDAI (The Graduate University for Advanced Studies), 2-21-1 Osawa, Mitaka, Tokyo 181-8588, Japan}
\affiliation{National Astronomical Observatory of Japan, 2-21-1 Osawa, Mitaka, Tokyo 181-8588, Japan}
\author[0000-0002-2364-0823]{Daisuke Iono}
\affiliation{Department of Astronomy, School of Science, SOKENDAI (The Graduate University for Advanced Studies), 2-21-1 Osawa, Mitaka, Tokyo 181-8588, Japan}
\affiliation{National Astronomical Observatory of Japan, 2-21-1 Osawa, Mitaka, Tokyo 181-8588, Japan}

\author[0000-0001-9728-8909]{Ken-ichi Tadaki}
\affiliation{Faculty of Engineering, Hokkai-Gakuen University, Toyohira-ku, Sapporo 062-8605, Japan}

\author[0000-0002-3560-8599]{Maximilien Franco}
\affiliation{Universit\'{e} Paris-Saclay, Universit\'{e} Paris Cit\'{e}, CEA, CNRS, AIM, 91191 Gif-sur-Yvette, France}

\author[0000-0001-7095-7543]{Min S. Yun}
\affiliation{University of Massachusetts Amherst 710 North Pleasant Street, Amherst, MA 01003-9305, USA}

\author[0000-0002-7051-1100]{Jorge A. Zavala}
\affiliation{University of Massachusetts Amherst 710 North Pleasant Street, Amherst, MA 01003-9305, USA}

\author[0000-0003-4807-8117]{Yoichi Tamura}
\affiliation{Institute for Advanced Research, Nagoya University, Furocho, Chikusa, Nagoya 464-8602, Japan}

\author[0000-0002-1499-6377]{Takafumi Tsukui}
\affiliation{Astronomical Institute, Tohoku University, 6-3, Aramaki, Aoba-ku, Sendai, Miyagi, 980-8578, Japan}

\author[0000-0003-2919-7495]{Christina C.\ Williams}
\affiliation{NSF National Optical-Infrared Astronomy Research Laboratory, 950 North Cherry Avenue, Tucson, AZ 85719, USA}

\author[0000-0001-6469-8725]{Bunyo Hatsukade}
\affiliation{Department of Astronomy, School of Science, SOKENDAI (The Graduate University for Advanced Studies), 2-21-1 Osawa, Mitaka, Tokyo 181-8588, Japan}
\affiliation{National Astronomical Observatory of Japan, 2-21-1 Osawa, Mitaka, Tokyo 181-8588, Japan}
%\affiliation{The Graduate University for Advanced Studies (SOKENDAI), Shonan Village, Hayama, Kanagawa 240-0193, Japan}
\affiliation{Department of Astronomy, Graduate School of Science, The University of Tokyo, 7-3-1 Hongo, Bunkyo-ku, Tokyo 133-0033, Japan}

\author[0000-0002-2419-3068]{Minju M. Lee}
\affiliation{Cosmic Dawn Center (DAWN), Denmark}
\affiliation{DTU Space, Technical University of Denmark, Elektrovej 327, DK2800 Kgs. Lyngby, Denmark}

\author[0000-0003-2475-7983]{Tomonari Michiyama}
\affiliation{Faculty of Information Science, Shunan University, 843-4-2, Gakuendai, Shunan, Yamaguchi 745-8566, Japan}

\author[0000-0001-7300-9450]{Ikki Mitsuhashi}
\affiliation{Department for Astrophysical \& Planetary Science, University of Colorado, Boulder, CO 80309, USA}

\author[0000-0002-6939-0372]{Kouichiro Nakanishi}
\affiliation{Department of Astronomy, School of Science, SOKENDAI (The Graduate University for Advanced Studies), 2-21-1 Osawa, Mitaka, Tokyo 181-8588, Japan}
\affiliation{National Astronomical Observatory of Japan, 2-21-1 Osawa, Mitaka, Tokyo 181-8588, Japan}

\author[0000-0002-0930-6466]{Caitlin M. Casey}
\affiliation{Cosmic Dawn Center (DAWN), Denmark}
\affiliation{Department of Physics, University of California, Santa Barbara, Santa Barbara, CA 93106, USA}

\author[0009-0008-9352-2938]{Soh Ikarashi}
\affiliation{Junior College, Fukuoka Institute of Technology, 3-30-1 Wajiro-higashi, Higashi-ku, Fukuoka 811-0295, Japan}

\author[0000-0003-4814-0101]{Kianhong Lee}
\affiliation{National Astronomical Observatory of Japan, 2-21-1 Osawa, Mitaka, Tokyo 181-8588, Japan}
\affiliation{Astronomical Institute, Tohoku University, 6-3, Aramaki, Aoba-ku, Sendai, Miyagi, 980-8578, Japan}
\affiliation{Department of Physics, Graduate School of Science, Nagoya University, Furocho, Chikusa, Nagoya 464-8602, Japan}

\author[0000-0003-1747-2891]{Yuichi Matsuda}
\affiliation{Department of Astronomy, School of Science, SOKENDAI (The Graduate University for Advanced Studies), 2-21-1 Osawa, Mitaka, Tokyo 181-8588, Japan}
\affiliation{National Astronomical Observatory of Japan, 2-21-1 Osawa, Mitaka, Tokyo 181-8588, Japan}

\author[0000-0002-2501-9328]{Toshiki Saito}
\affiliation{National Astronomical Observatory of Japan, 2-21-1 Osawa, Mitaka, Tokyo 181-8588, Japan}
\affiliation{Faculty of Global Interdisciplinary Science and Innovation, Shizuoka University, 836 Ohya, Suruga-ku, Shizuoka 422-8529, Japan}
\affiliation{Department of Science, Graduate School of Integrated Science and Technology, Shizuoka University, 836 Ohya, Suruga-ku, Shizuoka 422-8529, Japan}

\author[0000-0001-9500-604X]{Andrea Silva}
\affiliation{National Astronomical Observatory of Japan, 2-21-1 Osawa, Mitaka, Tokyo 181-8588, Japan}

\author[0000-0003-1937-0573]{Hideki Umehata}
\affiliation{Institute for Advanced Research, Nagoya University, Furocho, Chikusa, Nagoya 464-8602, Japan}
\affiliation{Department of Physics, Graduate School of Science, Nagoya University, Furocho, Chikusa, Nagoya 464-8602, Japan}

\author[0000-0002-1319-3433]{Hidenobu Yajima}
\affiliation{Center for Computational Sciences, University of Tsukuba, Ten-nodai, 1-1-1 Tsukuba, Ibaraki 305-8577, Japan}

\correspondingauthor{Ryota Ikeda}
\email{ryota195ikeda@gmail.com}

\begin{abstract}
We present an analysis of rest-frame optical and far-infrared continuum emission in three luminous submillimeter galaxies (SMGs) at $3.0\lesssim z\lesssim4.5$. The SMGs are spatially resolved down to 400-500\,pc ($\sim0\farcs05$) resolution by James Webb Space telescope (JWST) and Atacama Large Millimeter/submillimeter Array (ALMA) observations. Despite similarities in their observed far-infrared properties (flux density, infrared luminosity, and effective radius), the three SMGs exhibit  heterogeneous morphologies both across wavelengths and among the sources themselves. While two of them (AzTEC-4 and AzTEC-8) show a disk-like structure in optical continuum, AzTEC-1 is dominated by highly concentrated component {with the S\'{e}rsic index of $n=5.4$}, where its far-infrared continuum emission is clumpy and less concentrated. AzTEC-4, which is confirmed to be at $z=4.198$, shows a two-arm spiral of dust, but not in the stellar distribution. These three SMGs exemplify that multiple physical mechanisms exist in triggering starbursts in luminous SMGs at high redshift: secular instability in gas disks (AzTEC-4) in addition to possible minor mergers (AzTEC-8), and a combination of the efficient gas supply to the central core induced by a gas-rich major merger and the reformation of cold gas disk (AzTEC-1). 
\end{abstract}


\section{Introduction}
\label{sec:Section1}
Luminous submillimeter galaxies (SMGs) or dusty star-forming galaxies (DSFGs) \footnote{We use these two terms as their names imply; we refer to SMGs as galaxies that are bright in the submillimeter wavelengths ($S_{860\mu\mathrm{m}} \gtrsim 3$\,mJy), and DSFGs as galaxies that contain dust grains emitting FIR radiation and exhibit high star formation rates (SFR $\gtrsim 100\,M_\odot\,\mathrm{yr}^{-1}$).} in the distant Universe are thought to represent the progenitors of massive elliptical galaxies found in the centers of present-day galaxy clusters (e.g., \citealp{1999ApJ...515..518E}; \citealp{2006ApJS..163....1H};  \citealp{2014ApJ...782...68T}). {The stellar mass formed during the SMG phase constitutes a large fraction of the total stellar mass observed in the present-day Universe (\citealp{2010MNRAS.404.1775T}; \citealp{2011ApJ...735L..34G}), thus understanding the triggering mechanisms of active star formation in SMGs is crucial for uncovering how massive elliptical galaxies were formed.} Gas-rich major merger (\citealp{2010ApJ...722.1666W}; \citealp{2010ApJ...724..233E}) or minor mergers (\citealp{2011ApJ...735L..34G}; \citealp{2018ApJ...856..121G}), and large-scale gas inflow in isolated disks (\citealp{2009ApJ...703..785D}; \citealp{2015Natur.525..496N}) are the possible scenarios explaining the observed large star formation rates (SFRs). Hydrodynamical simulations predict that the SMG population is a mixture of merger-induced starbursts and isolated disks (\citealp{2013MNRAS.428.2529H}; \citealp{2019MNRAS.488.2440M}). However, due to the coarse angular resolution (lower than a kpc scale) of previous observations and the dust-obscured, optically faint nature of SMGs, the exact mechanisms triggering starbursts and the relative contributions of each mechanism remain observationally unconstrained.

Recently, the Near-Infrared Camera (NIRCam; \citealp{2023PASP..135b8001R}) on board the James Webb Space Telescope (JWST; \citealp{2023PASP..135f8001G}) has played a crucial role in characterizing the rest-frame near-infrared (NIR) morphologies of SMGs at $z\lesssim4$, as a tracer of stellar structure (e.g., \citealp{2022ApJ...939L...7C}; \citealp{2023ApJ...942L..19C}; \citealp{2023ApJ...958L..26H}; \citealp{2023ApJ...958...36S}; \citealp{2024A&A...691A.299G}; \citealp{2024A&A...690A.285P}; \citealp{2024ApJ...973...25K}; \citealp{2025ApJ...978..165H}; \citealp{2025ApJ...979..229M}; \citealp{2025ApJ...980...11P}; \citealp{2025PASJ...77..432U}; \citealp{2025ApJ...988..135M}). It has been claimed that the majority of SMGs exhibit smooth, disk-like stellar distributions with sub-structures (\citealp{2024A&A...690A.285P}; \citealp{2024A&A...691A.299G}; \citealp{2025PASJ...77..432U}) from the rest-frame NIR observations, but other studies claim that mergers and interactions are the dominant population in their sample \citep{2025ApJ...978..165H}. Merger fractions of starburst galaxies, based on visual classifications, are reported to increase from $z\sim1$ (a merger fraction of $\sim15$\,\%) to $z\sim3$ ($\sim27$\,\%; \citealp{2025ApJ...982..200R}; see also \citealp{2025ApJ...980..204F}), suggesting that mergers play a more significant role at higher redshift. 

\begin{deluxetable*}{cccccc}[ht!]
\tablecaption{Sample \label{tab:Table1}}
\tablewidth{0pt}
\tablehead{
\colhead{Galaxy} & \colhead{R.A.} & \colhead{Decl.} & \colhead{$z$} & \colhead{COSMOS2025 ID}& \colhead{other IDs\tablenotemark{$^{\rm a}$}} \\
\colhead{} & \colhead{(deg)} & \colhead{(deg)} & \colhead{} &
\colhead{}
}
%\decimalcolnumbers
\startdata
AzTEC-1 & 149.92855 & 2.49395 & 4.342 & 385499 & AzTEC/C5, AS2COS0023.1, eMORA.3 \\
AzTEC-4 & 149.88207 & 2.51223 & 4.198 \tablenotemark{$^{\rm b}$} & 385679 & AzTEC/C4, AS2COS0155.1 \\
AzTEC-8 & 149.99721 & 2.57805 & 3.097 & 400785 & AzTEC/C2, AS2COS0028.1 \\
AzTEC-8.2\tablenotemark{$^{\rm c}$} & 149.99795 & 2.57822 & $2.83^{+0.04}_{-0.03}$ & 400901 & AS2COS0028.2 \\
AzTEC-8.3\tablenotemark{$^{\rm c}$} & 149.99834 & 2.57657 & $3.00^{+0.03}_{-0.04}$ & 400614 & -- \\
\enddata
\tablecomments{ {\rm a.} \ The AzTEC/COSMOS ID (AzTEC/C) referred from \cite{2011MNRAS.415.3831A}. The ALMA-SCUBA2 COSMOS ID (AS2COS) referred from \cite{2020MNRAS.495.3409S}. The Extended MORA ID (eMORA) referred from \cite{2024arXiv240814546L}. \\ {\rm b.} \ Spectroscopically confirmed by the ALMA Band 3 observation (Section\,\ref{subsubsec:Section3.1.3}). \\ {\rm c.} \ Two serendipitously galaxies detected in ALMA Band 7 around AzTEC-8. The photometric redshifts are taken from the catalog of the COSMOS-Web Public Data Release\,1 \citep{2025arXiv250603243S}.}
\end{deluxetable*}

{Despite the recent breakthroughs enabled by the JWST NIRCam instrument, the images only provide information on stellar distribution and dust-unobscured star formation, which contribute only a small fraction to the total SFR in massive galaxies such as luminous SMGs (\citealp{2017ApJ...850..208W}; \citealp{2020ApJ...902..112B}; \citealp{2024ApJ...971..161M}).} In addition, several limitations remain in determining the origin of dusty starbursts at high redshifts. First, the NIRCam/F444W filter has the point spread function (PSF) sizes of full-width at half-maximum $({\rm FWHM})\sim0\farcs14$, and sub-structures finer than this scale cannot be resolved unless they are gravitationally lensed. While NIRCam filters with shorter wavelengths have sharper PSF size, the dust obscuration becomes more severe and the observed morphology would be biased toward less dust-obscured regions. Second, as SMGs are known to exist beyond $z=4$ and up to $z\sim7$ \citep{2018Natur.553...51M}, even the longest wavelengths in NIRCam (4.8\,$\mu$m) can only reach the bluer side of the rest-frame optical wavelengths ($\lambda\lesssim 0.6\,\mu{\rm m}$) for the most distant ones, where dust extinction can be significant. The JWST Mid-Infrared Instrument (MIRI; \citealp{2023PASP..135d8003W}), which operates at longer wavelengths ($\lambda\sim5-28$\,$\mu$m) can capture the stellar structure at these redshifts (\citealp{2023A&A...671A.105A}; \citealp{2023A&A...673L...6C}; \citealp{2024ApJ...969...27B}), but the spatial resolution becomes even worse (PSF ${\rm FWHM} \gtrsim 0\farcs2$), hampering the detailed characterization of the stellar component.

{An alternative and reliable approach to probing the sub-structures of distant SMGs is to investigate their detailed structure from kpc to sub-kpc scales through interferometric imaging of the far-infrared (FIR) emission lines using the Atacama Large Millimeter/submillimeter Array (ALMA). High-resolution gas kinematics using CO, [C\,{\sc i}], and [C\,{\sc ii}] lines has played a key role in disentangling whether SMGs are rotating disks or mergers, particularly when they are gravitationally lensed (e.g., \citealp{2015MNRAS.453L..26R}; \citealp{2020Natur.584..201R}; \citealp{2025MNRAS.536.3757A}; but also see  \citealp{2012ApJ...760...11H}; \citealp{2018ApJ...863...56C}; \citealp{2020ApJ...889..141T}; \citealp{2021Sci...371..713L}; \citealp{2023A&A...679A.129R}; \citealp{2025Natur.641..861H} for studies of unlensed galaxies). However, because of the high observational cost, it has been challenging to build a statistical sample of highly resolved SMGs using FIR emission lines redshifted to the ALMA bands.}

{On the other hand, FIR continuum emission, observed with ALMA Band 7 (observed-frame 860\,$\mu$m) for instance, is observationally more accessible than emission lines for achieving a high signal-to-noise ratio (S/N), and it can probe dust-obscured star formation at resolutions as fine as $0\farcs012$, corresponding to a physical scale of 80\,pc at $z=4$. Numerous works have leveraged the power of sub-kpc imaging of distant SMGs, uncovering the sub-structures such as clumps, bars, and spirals traced by cold dust emission (\citealp{2016ApJ...829L..10I}; \citealp{2018Natur.560..613T}; \citealp{2019ApJ...876..130H}; \citealp{2019ApJ...882..107R};  \citealp{2020MNRAS.494.5542R}; \citealp{2020MNRAS.495L...1I}; \citealp{2021Sci...372.1201T}; \citealp{2022ApJ...929L...3S}; \citealp{2024MNRAS.527.8941T}). The key question, however, is how these sub-structures are related to the origin of starbursts at high redshifts. Several works have combined sub-kpc FIR information with the JWST images (\citealp{2023ApJ...948L...8R}; \citealp{2025ApJ...978..165H}; \citealp{2025PASJ...77..432U}), but the interplay between sub-structures of interstellar medium (ISM), stellar morphology, and star formation properties are not fully understood, and clearly larger samples are necessary to draw general conclusions. Furthermore, {only a handful of luminous SMGs at $z>4$, which are intrinsically rare, have been thoroughly studied to date (e.g., \citealp{2015ApJ...798L..18H}; \citealp{2018Natur.560..613T}; \citealp{2018NatAs...2...56Z}; \citealp{2023MNRAS.521.1045R})}. These rare and extreme objects provide unique insights into the extreme conditions in the early Universe, making their study crucial for understanding the SMG population in general.}  

In this paper, we provide a detailed analysis toward three luminous SMGs: COSMOS-AzTEC-1, COSMOS-AzTEC-4, and COSMOS-AzTEC-8 (hereafter AzTEC-1, AzTEC-4, and AzTEC-8, respectively), as part of the Formation Of Sub-Structure In Luminous SMGs (FOSSILS; Ikeda et al. in preparation)\footnote{Initial sample of 12 FOSSILS sample is briefly described in \cite{2025arXiv250920720I}}. FOSSILS is an ALMA Band 7 survey of a large sample of luminous SMGs ($N\gtrsim30$), aiming to characterize their sub-structures by resolving the FIR continuum emission down to sub-kpc scales. All three SMGs studied in this paper are a sub-sample of the FOSSILS and are spatially resolved down to $\lesssim500$\,pc scales in both the rest-frame optical and FIR continua by JWST/NIRCam and ALMA observations. 

We first give a description of our sample and observations in Section \ref{sec:Section2}. The methods and our main results are given in Section \ref{sec:Section3}. In Section \ref{sec:Section4}, we dive into the origin of starbursts by comparing our sample with the SMGs studied in the literature and simulated galaxies.
%The summary of this paper will be given in Section \ref{sec:Section5}.
Throughout this paper, we assume a flat $\Lambda$CDM cosmology with $H_{0} = 70$\,km/s\,Mpc$^{-1}$, $\Omega_{\rm M}=0.3$, and $\Omega_{\Lambda}=0.7$. We adopt the Chabrier initial
mass function (IMF; \citealp{2003PASP..115..763C}) for the calculations of the stellar masses and SFRs.

\section{Sample and observations} 
\label{sec:Section2}

\subsection{Sample}
\label{subsec:Section2.1}

The sample of this study comprises three luminous SMGs in the Cosmic Evolution Survey (COSMOS) field \citep{2007ApJS..172....1S}. They were first reported by the 890\,$\mu$m survey of 1.1 mm-selected submillimeter sources in the COSMOS field (\citealp{2007ApJ...671.1531Y}; \citealp{2009ApJ...704..803Y}), using the AzTEC camera (\citealp{2008MNRAS.386..807W}; \citealp{2011MNRAS.415.3831A}) mounted on the James Clark Maxwell Telescope (JCMT). \cite{2016ApJ...829L..10I} presented high-resolution (0\farcs015 - 0\farcs05) images of the three brightest SMGs studied in this paper, which are verified to be single and unlensed sources, in 860\,$\mu$m continuum emission using ALMA Band 7. {They discovered multiple clumps (with sizes of $\sim200$ pc) extending over regions up to $\sim$3--4 kpc.} However, the lack of observations in short baselines causes the missing flux for more than half of the total flux according to that measured with the Submillimeter Array (SMA) observation. \cite{2025ApJ...979..168U} studied the presence of an active galactic nucleus (AGN) in SMGs in the COSMOS field based on X-ray detection and spectral energy distribution (SED) fitting analysis. None of the three SMGs show evidence for AGNs.

AzTEC-1 at $z_{\mathrm{spec}}=4.342$ \citep{2015MNRAS.454.3485Y} is the most luminous SMG among the parent sample presented in \cite{2007ApJ...671.1531Y}. High-resolution 860\,$\mu$m continuum imaging shows two off-centered submillimeter clumps \citep{2016ApJ...829L..10I}. Spatially-resolved CO $J=4-3$ and \ci ${^3}P_{1}-{^3}P_{0}$ line emission ($0\farcs08$-resolution) reveal a gravitationally unstable gas disk \citep{2018Natur.560..613T}. Furthermore, gas kinematics using spatially-resolved [C\,{\sc ii}]\,158$\mu$m line ($0\farcs17$-resolution) reveals two non-corotating gas clumps, in which one of them is cospatial to the submillimeter clump \citep{2020ApJ...889..141T}. These studies highlight the complex nature of AzTEC-1. 

High-resolution SMA and ALMA observations of AzTEC-4 and AzTEC-8 were reported by \cite{2010MNRAS.407.1268Y} and \cite{2016ApJ...829L..10I}, respectively. The $0\farcs05$-resolution image of AzTEC-4 consists of two sources that are separated by 1.5\,kpc, possibly indicating an on-going merger \citep{2016ApJ...829L..10I}. The spectroscopic redshift of AzTEC-4 has been confirmed to be $z_{\rm spec}=4.198$ by ALMA Band 3 spectral scan observations (Section\,\ref{subsubsec:Section3.1.3}).
%Only photometric redshift ($z_{\mathrm{phot}}=4.70^{+0.43}_{-1.11}$; \citealp{2012A&A...548A...4S}) is available for this source. 
AzTEC-8 is resolved into two clumps separated by $\sim200$\,pc in the central region \citep{2016ApJ...829L..10I}. \cite{2022ApJ...929..159C} report an ALMA Band 3 spectroscopic scan for this source, successfully determining the spectroscopic redshift of $z_{\mathrm{spec}}=3.097$. Finally, two additional galaxies (AzTEC-8.2 and AzTEC-8.3) around AzTEC-8 are serendipitously detected in ALMA Band 7 continuum (Section\,\ref{subsubsec:Section2.2.1}). While the spectroscopic redshifts of AzTEC-8.2 and AzTEC-8.3 are unconstrained to date, the photometric redshifts derived from the SED modeling are similar to the spectroscopic redshift of AzTEC-8. 

We summarize the coordinates, redshifts and other nomenclature of the galaxies studied in this paper in Table\,\ref{tab:Table1}.

\begin{figure*}[ht!]
\includegraphics[width=\linewidth]{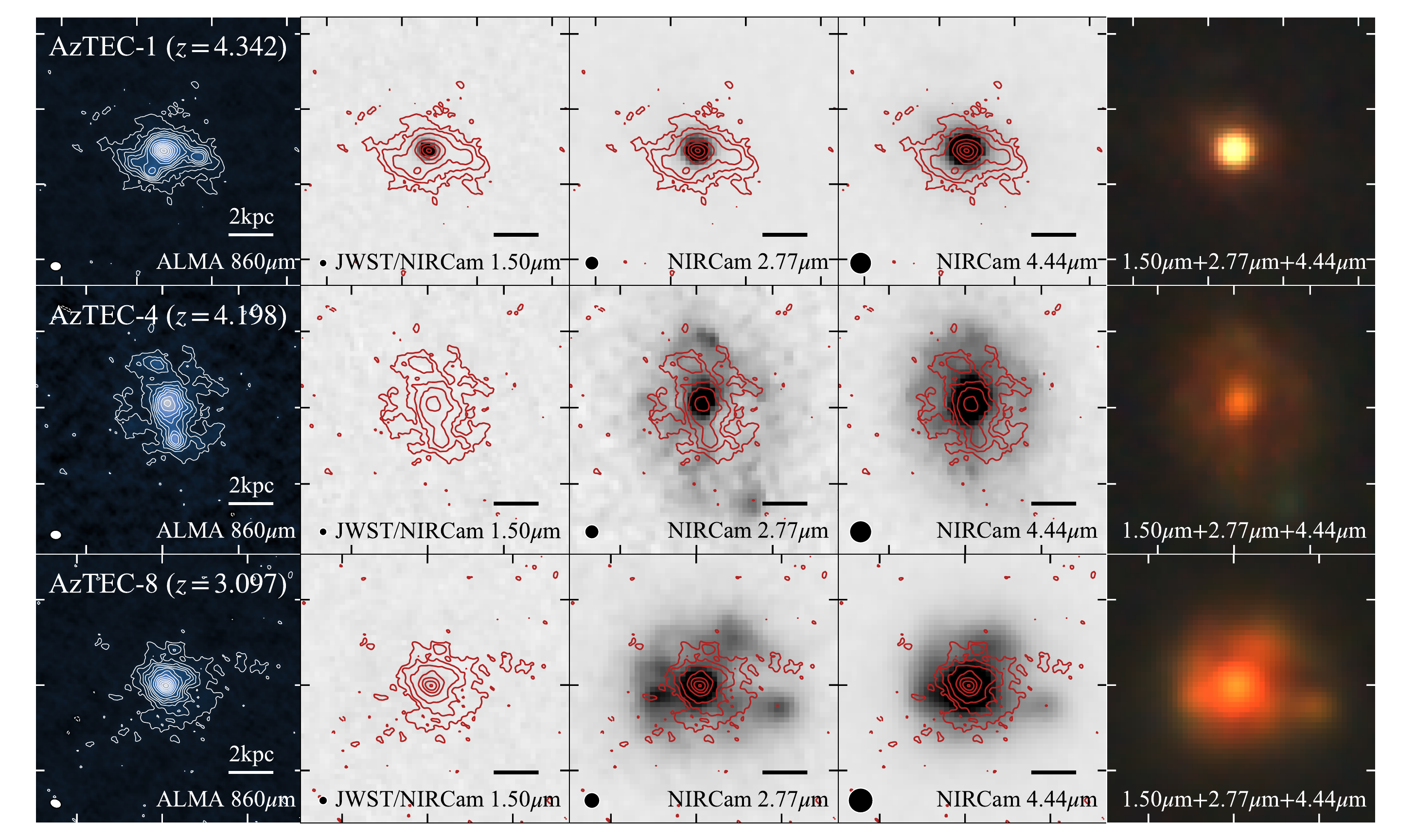}
\caption{Approximately $500$\,pc resolution ALMA and JWST images of three SMGs. From left to right, each panel shows ALMA 860\,$\mu$m (Band 7) continuum, JWST/NIRCam F150W, F277W, F444W, and RGB (F150W$+$F277W$+$F444W) images, covering a 12 kpc\,$\times$\,12 kpc area. We show the synthesized beam shape of ALMA and the PSF size of JWST images on the lower left corner as white and black ellipses, respectively. The white contours on the ALMA 860\,$\mu$m continuum images start at $2.5\sigma$ and increase in steps of $2.5\sigma$ until $15\sigma$ and $5\sigma$ until $50\sigma$. The contour levels of 860\,$\mu$m continuum on the JWST images are [2.5, 5, 10, 20, 30, 40, 50]\,$\sigma$. The tick spacing is $0\farcs5$. \label{fig:Figure1}}
\end{figure*}

\begin{deluxetable*}{ccccc}[t!]
\tablecaption{Summary of the ALMA data analyzed in this study \label{tab:Table2}}
\tablehead{
\colhead{Galaxy} & \colhead{Band} & \colhead{beam shape} & \colhead{{rms noise level}} & \colhead{spectral line} \\
\colhead{} & \colhead{} & \colhead{} & \colhead{{($\mu$Jy/beam)}} & \colhead{}
}
\startdata
\multirow{3}{*}{AzTEC-1} & 9 & $0\farcs19\times0\farcs15$, -$79.8^{\circ}$ & 264\,(continuum) & -- \\
 & 7 & $0\farcs077 \times 0\farcs062,\,87.3^{\circ}$ & 33.6\,(continuum) & -- \\
 & 4 & $0\farcs072\times 0\farcs059,\,32.9^{\circ}$ & 4.64\,(continuum) &  -- \\
\hline
\multirow{2}{*}{AzTEC-4} & 7 & $0\farcs077\times0\farcs062,\,87.1^{\circ}$ & 29.7\,(continuum) & -- \\
& 3 & $2\farcs28\times1\farcs93,\,58.5^{\circ}$ & 125\,(150\,km/s bin) & CO(4-3),\,[C\,{\sc i}](1-0),\,CO(5-4) \\
\hline
\multirow{3}{*}{AzTEC-8} & 7 & $0\farcs070\times0\farcs052,\,65.5^{\circ}$ & 31.4\,(continuum) & -- \\
& 6 & $0\farcs064\times0\farcs048,\,41.5^{\circ}$ & 31.6\,(continuum) & -- \\
& 3 &  $2\farcs72\times2\farcs37,\,79.4^{\circ}$ & 273\,(150\,km/s bin) & CO(3-2),\,CO(4-3) 
\enddata
\end{deluxetable*}

\subsection{Observations}

In this paper, we mainly analyzed ALMA (Bands 3, 4, 6, and 7) and JWST/NIRCam data. All of the ALMA data used in this paper were calibrated using Common Astronomy Software Application package ({\tt CASA}; \citealp{2022PASP..134k4501C}) and the ALMA Pipeline \citep{2023PASP..135g4501H} with the versions described in the Quality Assurance Level 2 (QA2) report .
We briefly summarize the ALMA data analyzed in this paper in Table\,\ref{tab:Table2}.

\subsubsection{ALMA Band 7 ($\lambda=860\,\mu m$)}
\label{subsubsec:Section2.2.1}

% AzTEC-1,4,8 (ext): 2015.1.01345.S (PI:Iono)
% AzTEC-1,4 (com): 2017.1.00127.S (PI:Iono)
% AzTEC-8 (com): 2016.1.00012.S (PI:Iono)
In order to reconstruct flux-complete high-resolution images of submillilimeter continuum, we combine observational data from three different ALMA programs, which cover different {spatial scales}. The extended configuration data, which achieves the highest angular resolution of $0\farcs015$ as presented in \cite{2016ApJ...829L..10I}, were initially taken as part of the ALMA Cycle 3 program (\#\,2015.1.01345.S) with a $uv$ range of $750\,{\rm k}\lambda \lesssim uv \lesssim 7500\,{\rm k}\lambda$. Subsequently, the compact configuration data were taken in Cycle 4 (\#\,2016.1.00012.S; $uv \lesssim 700\,{\rm k}\lambda$) for AzTEC-8 and in Cycle 5 (\#\,2017.1.00127.S; $uv \lesssim 1200\,{\rm k}\lambda$) for AzTEC-1 and AzTEC-4.

We flagged three bad antennas (DA45, DV03, and DV10) in Cycle\,3 data which cause artificial ripples in the reconstructed images. 
Then, we combine visibility data by using the {\tt CASA/concat} task. As the visibility weights taken in different programs in multiple ALMA cycles are not fine-tuned to create a {synthesized} Gaussian beam, we applied {\tt CASA/statwt} to put the relative weight on different configuration data. % Need for re-imaging AzTEC-1 Band 7 w/o contsub?
%For AzTEC-1, we applied {\tt CASA/uvcontsub} to model the underlying continuum emission without the frequency range corresponding to the redshifted \cii line emission. Other emission line-free spectral windows are fully used for the continuum fitting.

We apply the multi-scale \citep{2008ISTSP...2..793C} and auto-multithresh CLEAN algorithm. We use a 0\farcs02 $uv$-taper with the Briggs weighting (a robust parameter $R=0.5$). For the auto-multithresh CLEAN algorithm, the automasking parameters of $\rm noisethreshold=4.0$, $\rm lownoisethreshold=1.5$, and $\rm minbeamfrac=0.3$ are used for all SMGs. For $\rm sidelobethreshold$, we use 2.5 for AzTEC-1, and 2.0 for AzTEC-4 and AzTEC-8, which are determined by visually inspecting the spatial coverage of the created CLEAN mask on the SMGs. All CLEAN masks are CLEANed down to $1.5\sigma$ significance level. The final synthesized beam sizes are $0\farcs077$ $\times$ $0\farcs062$ with a position angle (PA) of ${\rm PA}=87.3^{\circ}$ for AzTEC-1 and AzTEC-4, and $0\farcs070$ $\times$ $0\farcs052$ with ${\rm PA}=65.5^{\circ}$ for AzTEC-8, corresponding to $\sim$460\,pc resolution. The root mean square (rms) noise levels range $29-34$\,$\mu$Jy/beam.

\begin{figure*}[t!]
\centering
\includegraphics[width=0.85\linewidth]{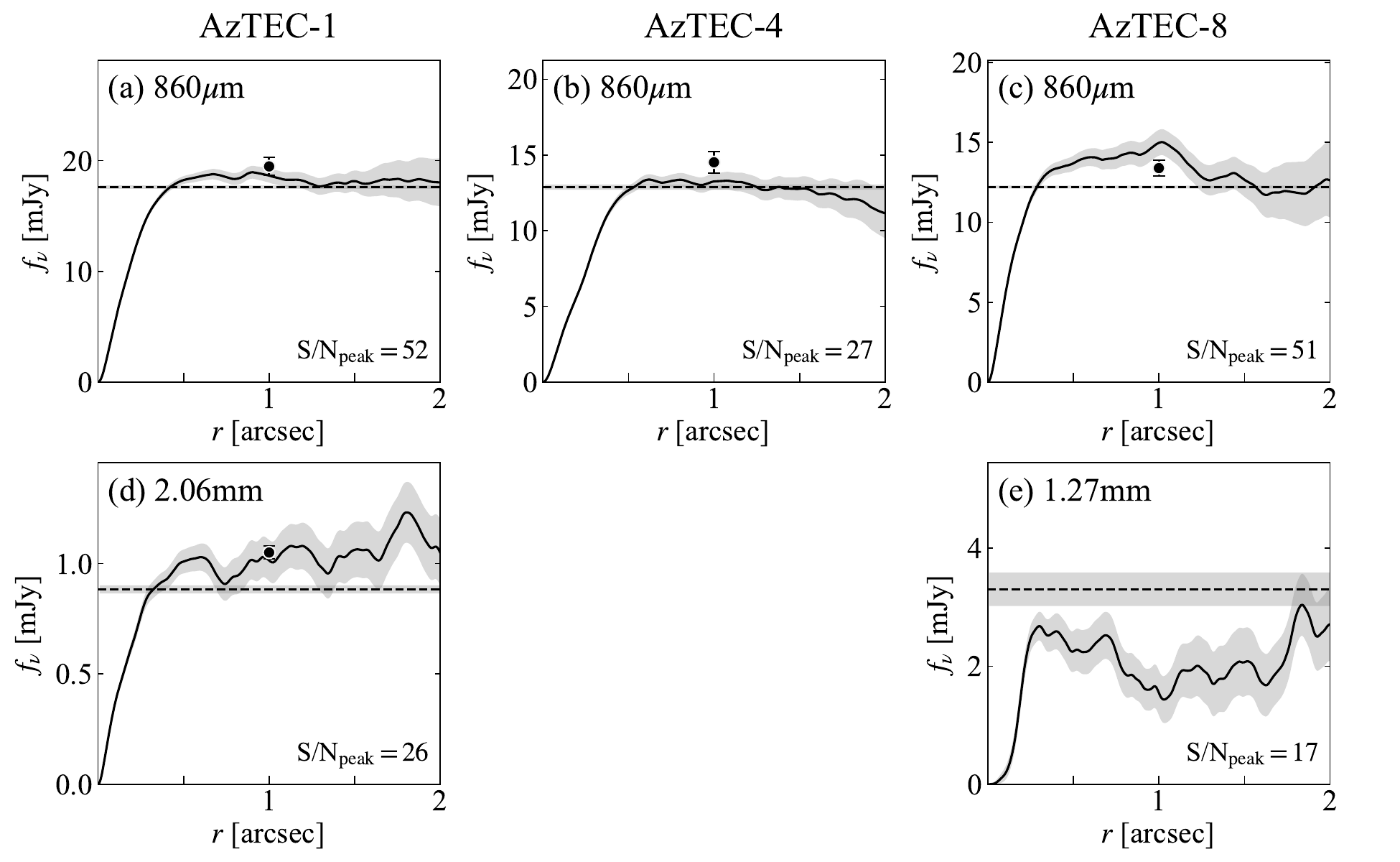}
\caption{Curve of growth of the flux density of the 860\,$\mu$m (panel a, b, and c), 2.06\,$\mu$m (panel d for AzTEC-1 only), and 1.27\,$\mu$m (panel e for AzTEC-8 only) continuum emission. The dashed lines are the total flux estimated from modeling the visibility data. The uncertainties are estimated by calculating the standard deviation of the growth curves obtained from randomly placing apertures on a blank sky region. The black circle at $r=1''$ indicates the aperture photometry, measured from the low-resolution images when available. The peak S/N of each image is shown in the lower-right corner of the corresponding panel. \label{fig:Figure2}}
\end{figure*}

The ALMA Band 7 images of the three SMGs are shown in the leftmost panels of Figure\,\ref{fig:Figure1}. To verify whether the total flux is recovered, we compare the curve of growths of the flux density as a function of radius, centered at the peak position of the CLEAN images, with the flux density inferred from the extrapolated zero-baseline amplitude by modeling the visibility in Figure\,\ref{fig:Figure2} (panels a, c, and d). The visibility modeling simultaneously provides the effective radius of 860\,$\mu$m emission as we will discuss in Section\,\ref{subsec:Section3.2}. We used {\tt UVMULTIFIT} \citep{2014A&A...563A.136M} by applying an elliptical exponential disk model. In addition, we plot the flux density measured from $2''$-aperture using the CLEAN image created from the data taken with the compact array configuration only ($0\farcs2$-resolution). The uncertainties are measured by setting random apertures on a blank sky region and then computing the standard deviation of the fluxes within the apertures.\footnote{The rms noise level measured using the random aperture method has been verified to be consistent with the covariance-based estimate that accounts for correlated noise in the interferometric image \citep{2023JATIS...9a8001T}.} Overall, the total fluxes in $\sim0\farcs06$-resolution images fully recover those in $\sim0\farcs2$-resolution images.

\begin{figure}[hb!]
\center
\includegraphics[width=0.88\linewidth]{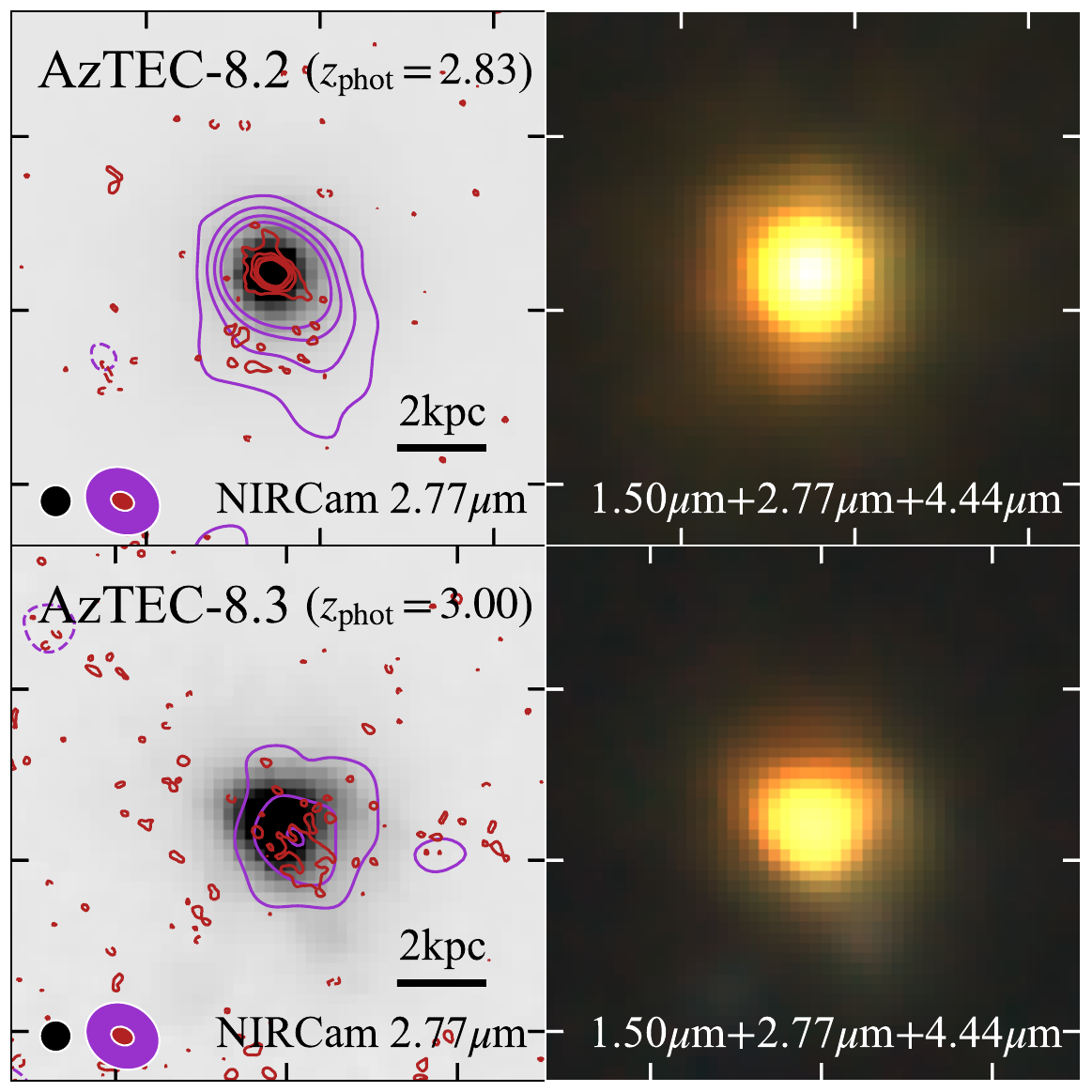}
\caption{Left: Comparison of ALMA 860\,$\mu$m continuum emission (contours) and NIRCam F277W images of AzTEC-8.2 and AzTEC-8.3. Two types of resolution (red: $0\farcs06$, purple: $0\farcs2$) are used to show the ALMA contours to demonstrate that the detection is significant ($>7\sigma$) in the low-resolution image. The contour levels both increase in steps of $2.5\sigma$ until $10\sigma$. Right: The RGB images (F150W$+$F277W$+$F444W). The tick spacing is $0\farcs5$. \label{fig:Figure3}}
\end{figure}

Two galaxies, AzTEC-8.2 and AzTEC-8.3, are serendipitously detected in the ALMA Band 7 image of AzTEC-8 with a flux density of 1.82\,mJy and 0.37\,mJy, respectively. Figure\,\ref{fig:Figure3} shows the ALMA Band 7 contours of the two galaxies overlaid on the NIRCam F277W image (Section\,\ref{subsubsec:Section2.2.5}). Since the detection of AzTEC-8.3 is relatively weak ($\sim3\sigma$) in the high-resolution image, we also present contours from the 0$\farcs$2-resolution image, created using the compact configuration data only. 

\subsubsection{ALMA Band 6 ($\lambda=1.3\,mm$)}
% 2017.1.00487.S (PI: Kristensen)

\begin{figure*}[ht!]
\center
\includegraphics[width=0.6\linewidth]{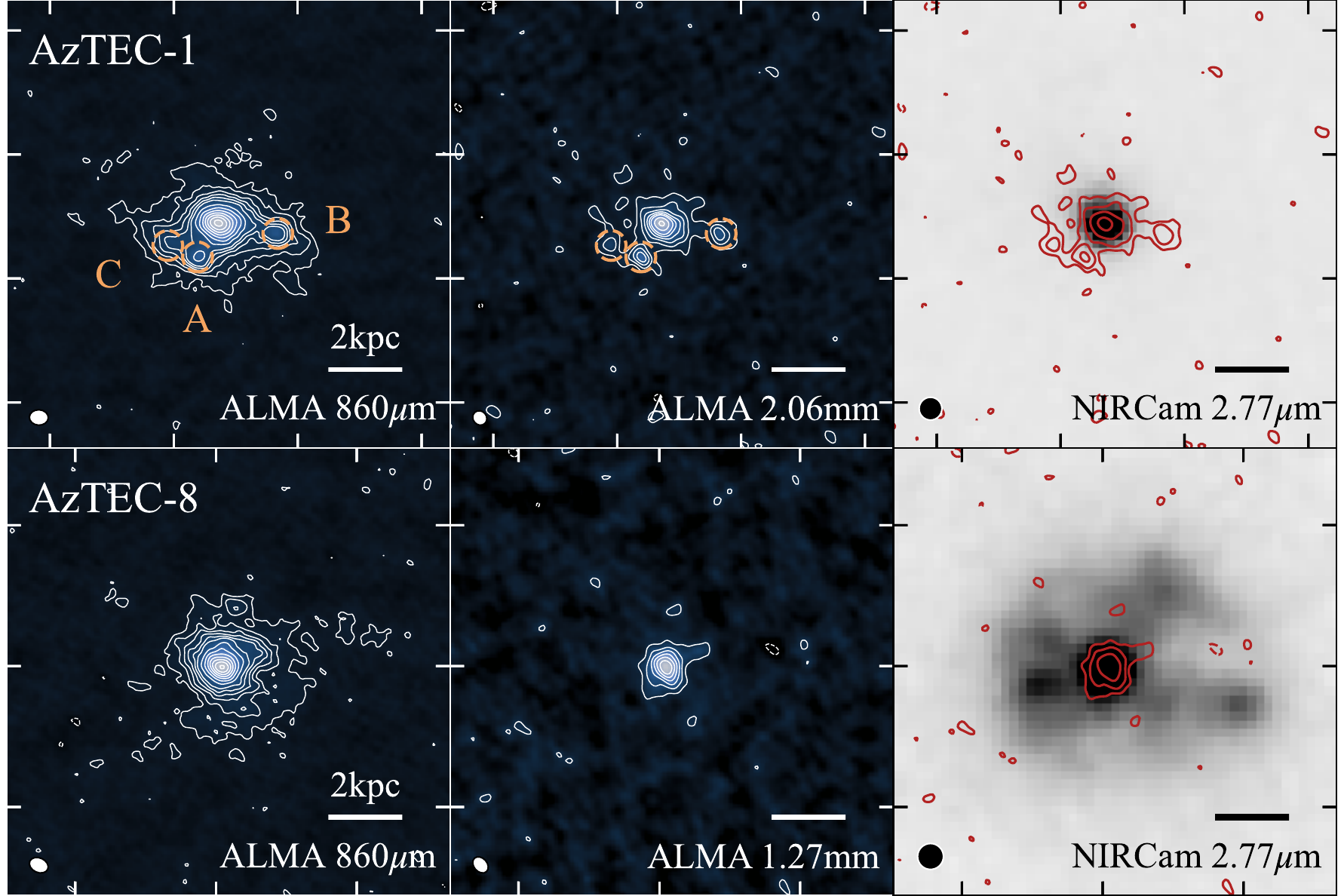}
\caption{Comparison of high resolution multi-band ALMA images. The top panels compare ALMA 860\,$\mu$m (Band 7) and 2.06\,mm (Band 4) continua of AzTEC-1, and the bottom panels compare ALMA 860\,$\mu$m and 1.27\,mm (Band 6) continua of AzTEC-8. The rightmost panels simultaneously show NIRCam F277W and ALMA Band 4/6. Three FIR clumps (A, B, and C) in AzTEC-1 are highlighted in orange dashed circles, which are the $0\farcs12$-aperture that we used for measuring fluxes. The contour levels and the tick spacing are the same as Figure\,\ref{fig:Figure1}.\label{fig:Figure4}}
\end{figure*}

AzTEC-8 was observed in the Band 6 (\#\,2017.1.00487.S) using the extended array configuration ($480\,{\rm k}\lambda \lesssim uv \lesssim 5500\,{\rm k}\lambda$). We follow the same procedures for imaging as Band 7, but applying {the Briggs weighting with $R=2.0$} to increase the S/N and to recover the extended emission. We excluded three bad antennas (DA06, DV09, and DV61) which cause the ripples in the image. The observations were performed using two spectral setups, covering $\sim4$\,GHz range centered at 221.6, 236.8, 249.3, and 263.9 GHz ($\sim 16$\,GHz frequency width in total). From the dirty cube images, we do not find any emission line at the position of neither AzTEC-8 nor the two Band 7 continuum sources, and hence the spectroscopic redshifts of AzTEC-8.2 and AzTEC-8.3 are unconfirmed. We apply a $0\farcs03$ $uv$-taper to the Band 6 data so that the spatial resolution is consistent with the Band 7 images. The final synthesized beam size is $0\farcs064\times 0\farcs048$ with ${\rm PA}=41.5^{\circ}$. The rms noise level is 31.6\,$\mu$Jy/beam. We show the curve of growth of 1.3\,mm continuum emission of AzTEC-8 in panel (e) of  Figure\,\ref{fig:Figure2}. Compared to the total flux ($3.3 \pm 0.3$\,mJy) derived from the visibility modeling, the flux is systematically lower by $\sim60$\,\%, indicating that some extended flux is lost, presumably due to the lack of short baseline data. The ALMA Band 6 image of AzTEC-8 is presented in Figure\,\ref{fig:Figure4}.

\subsubsection{ALMA Band 4 ($\lambda=2.1\,mm$)}
% AzTEC-1: 2018.1.01103.S

AzTEC-1 was observed in Band 4, using multiple configurations through the Cycle 6 program  (\#2018.1.01136.S; $70\,{\rm k}\lambda \lesssim uv \lesssim3600\,{\rm k}\lambda$). Two redshifted emission lines (CO\,$J=7-6$ and [C\,{\sc i}] $^{3}P_{2}-^{3}P_{1}$ lines) fall in one of the four spectral windows, thus we use the remaining three spectral windows to image the continuum emission. We follow the same approach as Band 7 imaging to create the CLEAN image, but modified the automasking parameters as $\rm noisethreshold = 3.5$ and $\rm sidelobethreshold = 1.5$, and applied a $0\farcs03$ $uv$-taper. The final synthesized beam size and the rms noise level are $0\farcs061\times0\farcs055$ with ${\rm PA}=40.9^{\circ}$ and 4.64\,$\mu$Jy/beam, respectively. We show the curve of growth of 2.1\,mm continuum emission of AzTEC-1 in panel (b) of  Figure\,\ref{fig:Figure2}. Compared to the total flux ($0.88 \pm 0.02$\,mJy) derived from the visibility modeling, the cumulative flux at $1''$ radius is higher by $\sim15$\,\%, but consistent with the photometry from the $0\farcs2$-resolution image. The ALMA Band 4 image of AzTEC-1 is presented in Figure\,\ref{fig:Figure4}.

\subsubsection{ALMA Band 3 ($\lambda=3.0\,mm$)}
% AzTEC-4: 2017.A.00034.S (PI: Iono) 
% AzTEC-8: 2019.1.01600.S (PI: Chen) 

In order to study the gas and dynamical properties traced by CO and \ci lines, we analyzed the data taken by two ALMA Band 3 spectral scan observations, targeting AzTEC-4 (\#\,2017.A.00034.S) and AzTEC-8 (\#\,2019.1.01600.S).
While the spectroscopic analysis of AzTEC-8 has already been reported by \cite{2022ApJ...929..159C} and \cite{2024ApJ...961..226L}, we revisit the data to estimate the dynamical mass (Section\,\ref{subsubsec:Section3.1.2}). The CO and \ci line properties of AzTEC-1 are reported in \cite{2018Natur.560..613T}.

Both observations cover  continuous frequency range of Band 3 ($\sim84-116$\,GHz) consisting of five spectral setups. 

\begin{figure*}
\centering
\includegraphics[width=0.95\linewidth]{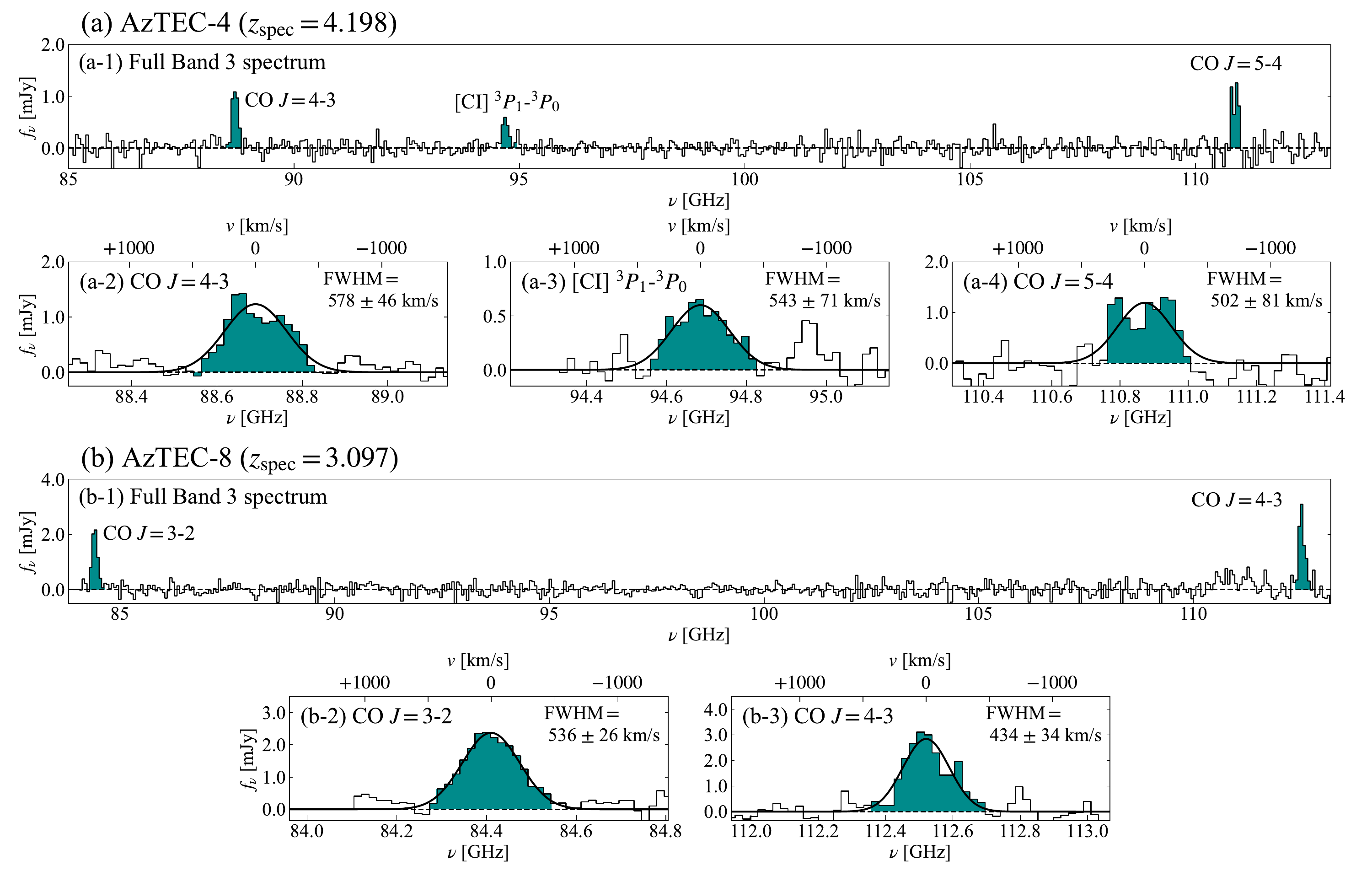}
\caption{Continuum-subtracted ALMA Band 3 spectra of (a) AzTEC-4 and (b) AzTEC-8. Both full spectral scan with $\sim150 - 200$\,km/s resolution (panels a-1 and b-1) and the close-ups of each emission line with 60\,km/s resolution (panels a-2, a-3, a-4, b-2, and b-3) are shown. All lines are fitted in a single Gaussian (solid curve). FWHM of the best-fit Gaussian is shown at the top right corner of each panel.\label{fig:Figure5}}
\end{figure*}
We made two images with different spectral resolution: using a full spectrum with $\sim150-200$\,km/s resolution and a partial spectrum focused on each emission line with 60\,km/s resolution. We CLEANed the dirty cubes down to $1.5\sigma$ using Briggs weighting ($R=0.5$). 
The beam size and rms noise level are $\sim2\farcs3$ and 125 $\mu$Jy/beam per 150\,km/s bin for AzTEC-4, and $\sim2\farcs5$ and 273 $\mu$Jy/beam per 150\,km/s bin for AzTEC-8. The Band 3 spectra of AzTEC-4 and AzTEC-8 are shown in Figure\,\ref{fig:Figure5}. As reported in \cite{2022ApJ...929..159C}, no emission lines are detected at the position of AzTEC-8.2 and AzTEC-8.3, most likely due to the lack of sensitivity.

\subsubsection{JWST NIRCam}
\label{subsubsec:Section2.2.5}
Three SMGs were observed in four JWST NIRCam filters (F115W, F150W, F277W, and F444W) through COSMOS-Web, a JWST Cycle 1 treasury survey program \citep{2023ApJ...954...31C}. The COSMOS-Web NIRCam dataset was processed using the JWST Calibration Pipeline \citep{2024zndo..10870758B}, with additional steps implemented to enhance image fidelity and astrometric accuracy. A complete overview of the data reduction process will be detailed in \cite{2025arXiv250603256F}, while we provide a brief summary here. The raw exposures from NIRCam were retrieved from the Mikulski Archive for Space Telescopes (MAST) and reduced using pipeline version 1.14.0. Several supplementary corrections were applied inspired by \cite{2024ApJ...965L...6B}, including mitigation of $1/f$ noise, background modeling, artifact removal, and masking of defective pixels. The calibration employed reference files from the Calibration Reference Data System (CRDS) pmap-1223, corresponding to NIRCam instrument mapping imap-0285. The final mosaics were constructed at a pixel scale of $0\farcs03$/pixel, balancing spatial resolution and photometric accuracy. The PSF for each NIRCam filter was reconstructed using {\tt PSFEx} \citep{2011ASPC..442..435B}.

To refine the astrometry, we used the JWST/Hubble Space Telescope (HST) Alignment Tool (JHAT; \citealp{2023zndo...7892935R}) to align the NIRCam images with a reference catalog derived from HST/ACS F814W mosaics \citep{2007ApJS..172..196K}. The alignment was further adjusted using Gaia Early Data Release 3 (EDR3; \citealp{2021A&A...649A...1G}), achieving sub-5 mas median absolute positional offsets and a median absolute deviation (MAD) below 12 mas across all bands.

The F150W, F277W, F444W, and RGB (F150W$+$F277W$+$F444W) images of three SMGs are shown in Figure\,\ref{fig:Figure1}. The F277W and RGB images of AzTEC-8.2 and AzTEC-8.3 are shown in Figure\,\ref{fig:Figure3}. To create the RGB images, we use {\tt pypher} \citep{2016zndo.....61392B} to match the PSF of F150W and F277W to that of F444W.

\begin{figure*}[ht!]
\center
\includegraphics[width=0.85\linewidth]{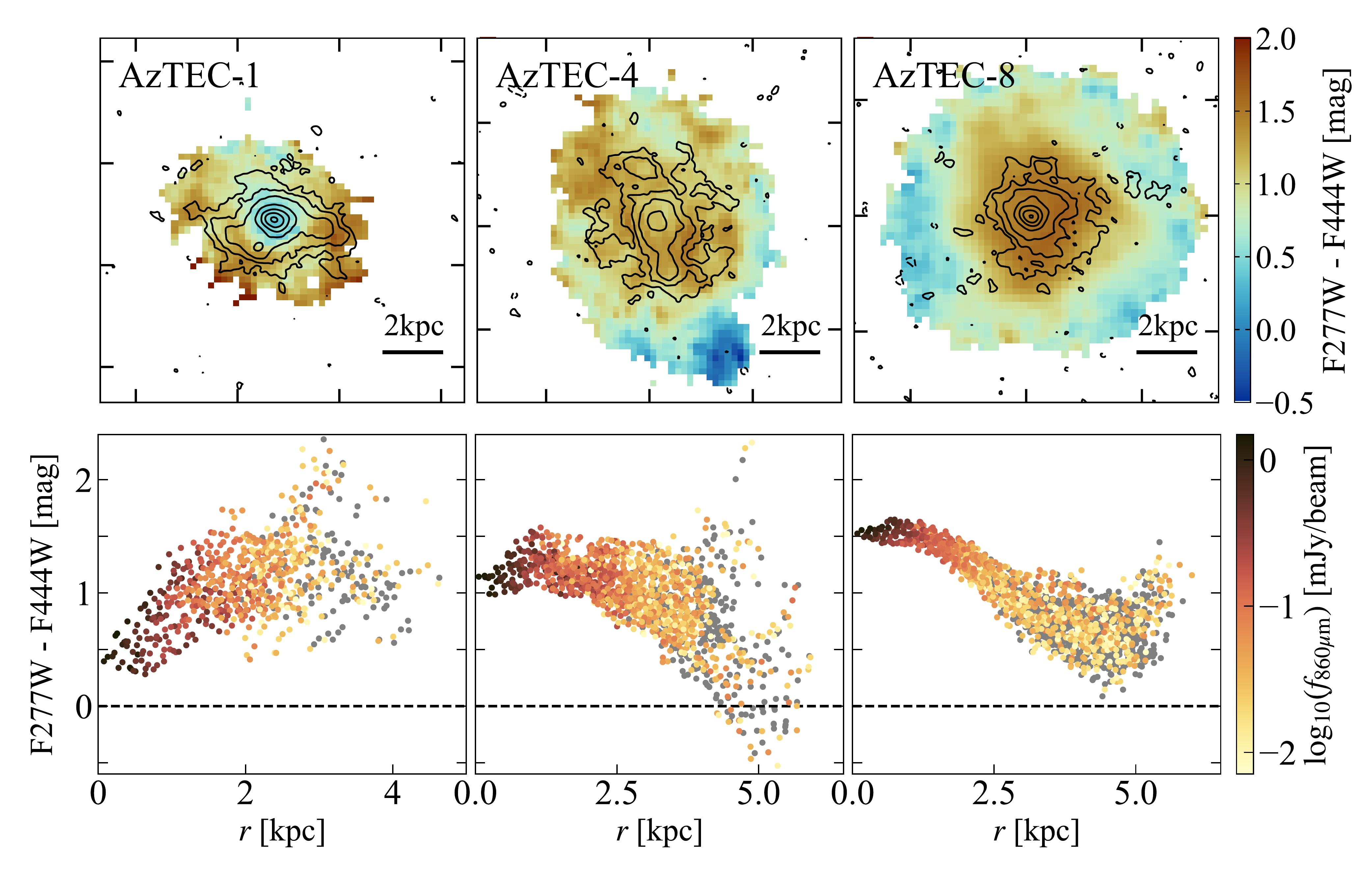}
\caption{Top: F277W - F444W color map of three SMGs shown {in units of magnitude}. The pixels detected below $4\sigma$ in both filters are clipped. The ALMA 860\,$\mu$m continuum image is overlapped (black contours) with the contour levels of [2.5, 5, 10, 20, 30, 40, 50]\,$\sigma$. The tick spacing is $0\farcs5$. Bottom: Radial profiles of F277W - F444W color centered at the peak position of 860\,$\mu$m continuum emission. Each datapoint corresponds to a pixel in JWST image and  color-coded by the 860\,$\mu$m continuum flux density. The grey datapoints signify where 860\,$\mu$m continuum emission is below the $2.5\sigma$ significance level.\label{fig:Figure6}}
\end{figure*}

\section{Results}
\label{sec:Section3}

These three SMGs are representative of the most FIR-luminous and intense starbursts during the first two billion years of the Universe, yet they exhibit few similarities in the morphology when high-resolution, multi-wavelength images are compared together. Figure\,\ref{fig:Figure1} shows a combined view of the rest-frame ultraviolet (UV), optical, and FIR continuum emission of the three SMGs. From UV to optical wavelengths, AzTEC-1 shows a concentrated morphology, whereas dust-obscured star formation takes place in the more extended region. AzTEC-4, which has previously been suggested as a possible merger in \cite{2016ApJ...829L..10I}, shows a smooth and extended morphology in both optical and in FIR continuum emission. The FIR continuum emission shows a spiral-like morphology, which we will explore in Section\,\ref{subsec:Section3.2}. In contrast to AzTEC-1, the optical morphology of AzTEC-8 is more complex, exhibiting more extended and clumpy features than the FIR continuum emission. Both AzTEC-4 and AzTEC-8 are not detected in UV wavelengths (F115W and F150W filters), suggesting dust obscuration throughout the disk. The compactness of the FIR continuum in these two galaxies (circularized radii of $R_e = 0.63$--$1.16$\,kpc; Section\,\ref{subsec:Section3.2}) is broadly consistent with the interpretation that dust obscuration in so-called optically dark galaxies (defined as e.g., $H>$27\,mag;  \citealp{2019Natur.572..211W}) is due to their compact FIR sizes (median $R_{e} = 1.00$\,kpc), as compared to optically bright galaxies (median $R_{e}=1.17$\,kpc; \citealp{2021MNRAS.502.3426S}).

Figure\,\ref{fig:Figure4} presents the comparisons of $\sim500$\,pc-resolution FIR images of AzTEC-1 and AzTEC-8, observed in different ALMA bands. The 2.06 mm image of AzTEC-1 exhibits three off-center clumps detected at $>5\sigma$ significance which are all spatially consistent with the three clumps seen in the 860\,$\mu$m image. The nature of the two brightest clumps were discussed in \cite{2016ApJ...829L..10I}, \cite{2018Natur.560..613T}, and \cite{2020ApJ...889..141T}, referred to as `clump-2' and `clump-3' (hereafter clump A and B; Figure\,\ref{fig:Figure4}). The clump located east of the main component (hereafter clump C) has not been reported. This is likely because either the sensitivity was insufficient or clump C was resolved out in the $0\farcs02$ image used in \cite{2016ApJ...829L..10I}. On the basis of the detection in multiple ALMA bands, we conclude that those three FIR clumps are real. In contrast, the optical counterpart of these FIR clumps are not detected in the 2.77$\mu$m image, implying that these are young and massive clumps in the formation phase \citep{2020ApJ...889..141T}. We measure the flux density of each clump using an aperture with $0\farcs12$ diameter. Clump A has a flux density of  $1.15\pm0.07$\,mJy at 860\,$\mu$m and $0.081\pm0.015$\,mJy at 2.06\,mm, clump B has $0.93\pm0.07$\,mJy at 860\,$\mu$m and $0.068\pm0.015$\,mJy at 2.06\,mm, and clump C has $0.76\pm0.08$\,mJy at 860\,$\mu$m and $0.048\pm0.015$\,mJy at 2.06\,mm, where the uncertainties are derived from setting the random aperture on a blank sky region. The sum of the three clumps accounts for $\sim16$\,\% and $\sim22$\,\% of the total flux density in 860\,$\mu$m and 2.06\,mm continua, respectively.

The 1.27\,mm continuum of AzTEC-8 is compact ($R_{e,\rm major}=0.61\pm0.06$\,kpc) and no prominent sub-structure is detected in both 860\,$\mu$m and 1.27\,mm continuum images. This compact component is cospatial with the central stellar core seen in the 2.77\,$\mu$m image, suggesting an intense star formation activity. However, given that the fraction of recovered flux of 1.27\,mm is $\sim50-60$\,\%  (Figure\,\ref{fig:Figure2}) and the rest-frame UV continuum observed by the NIRCam/F150W is undetected (Figure\,\ref{fig:Figure1}), it is possible that the true structure is more extended than the 1.27\,mm image presented here, suggesting the presence of widespread dust obscuration. We also note that the central component of 860\,$\mu$m continuum in AzTEC-8 splits into two clumps separated by $0\farcs025$ (200\,pc) when observed in $\sim0\farcs02$ resolution \citep{2016ApJ...829L..10I}, which are unresolved both in ALMA and JWST images shown in Figure\,\ref{fig:Figure1}. We defer the detailed characterizations of the FIR clumps in a future work with a larger sample.

Figure\,\ref{fig:Figure6} shows the spatial variation and the radial profile of the F277W - F444W color of the three SMGs, constructed by matching the PSFs of the F277W and F444W images using {\tt pypher}. As the spatial resolution is smoothed to the PSF of F444W (FWHM$=0\farcs14$), the clumpy structure seen in the F277W image of AzTEC-8 is smoothed out in the color map. The blue region seen $\sim1\arcsec$ southwest of the center of AzTEC-4 is more likely a less obscured part of the extended star-forming disk, rather than a satellite galaxy, since the redshifted H$\beta$ and [O\,{\sc iii}]\,5007\AA\, emission lines fall within the wavelength coverage of F277W and may boost the flux density in the F277W filter.
%In AzTEC-4, we can see that a blue region exists in south to the main component, which is likely to be a less obscured region as 860\,$\mu$m continuum emission is weak. 
The bottom panels of Figure\,\ref{fig:Figure6} reveals different characteristics of the radial color profile among the three SMGs. AzTEC-1 shows a positive gradient as a function of radius, i.e., the larger the radius, the redder it gets, whereas AzTEC-4 and AzTEC-8 shows a decreasing trend with radius, {i.e., the larger the radius, the bluer it gets.} The color gradient can be influenced by multiple factors (stellar age, metallicity, AGN, and dust reddening).  Since the UV continuum emission in AzTEC-4 and AzTEC-8 is undetected due to dust obscuration, reddening can naturally explain the negative color gradients seen in these sources. The positive color gradient observed in AzTEC-1 is attributed to its more compact morphology in F277W compared to F444W, indicating that its dust obscuration properties may differ from those of AzTEC-4 and AzTEC-8. The global values of the infrared excess, $\log_{10}({\rm IRX}=L_{\rm IR}/L_{\rm UV})=1.3$, and the UV slope, $\beta_{\rm UV}=-0.3$, place AzTEC-1 consistently along the canonical IRX--$\beta$ relation \citep{1999ApJ...521...64M}, suggesting that a homogeneous mixture of dust and stars, rather than a uniform dust screen, is preferred for AzTEC-1 (e.g., \citealp{2017MNRAS.472.2315P}).

In the following subsections, we aim to dissect the nature of three SMGs using two approaches. In Section\,\ref{subsec:Section3.1}, we characterize the global properties to put these SMGs into the broader context of high-redshift galaxies. In Section\,\ref{subsec:Section3.2}, we present morphological analyses on both ALMA and JWST images.

\subsection{Global properties}
\label{subsec:Section3.1}

\subsubsection{Stellar mass and star formation rate}
\label{subsubsec:Section3.1.1}

The galaxy-integrated stellar mass and SFR are two fundamental properties that describe galaxies within the framework of stellar mass buildup. In this study, we adopt the stellar mass and SFR derived by \cite{2025ApJ...979..229M}, who modeled $\sim300$ submillimeter detected galaxies in the COSMOS-Web field, including the three SMGs studied in this paper, by using the {\tt CIGALE} SED fitting code \citep{2019A&A...622A.103B}. 
The SED fitting was performed using ALMA archival data as well as an extensive photometric catalog from COSMOS2020 \citep{2022ApJS..258...11W}, assuming a combined star formation history of a delayed-$\tau$ model and a recent ($50$--$300$\,Myr) burst, with a Chabrier IMF. About 90\,\% of the ALMA archival data comes from either Band\,6 or Band\,7. For more details on the photometry and SED fitting, we refer the readers to \cite{2025ApJ...979..229M}. We list the SED-based stellar masses and SFRs of the three SMGs in Table\,\ref{tab:Table3}.

\subsubsection{Infrared luminosity}
\label{subsubsec:Section3.1.2}

As most or all of the rest-frame UV continuum is attenuated due to dust obscuration (Figure\,\ref{fig:Figure1}), the bulk of the star formation activity can be quantified by dust-obscured SFR. We estimate the IR luminosity as a representation of the dust-obscured SFR in the three SMGs.

In order to estimate the IR luminosity ($L_{\rm IR}$), an integrated luminosity at the wavelength range of 8-1000\,$\mu$m, it is crucial to constrain the peak of the FIR SED (rest-frame 50-200$\mu$m). In addition to Band 4 and 7 observations, AzTEC-1 has been observed in Band 9 ($\lambda=470$\,$\mu$m) at $\lesssim0\farcs5$ resolution \citep{2019ApJ...876....1T}, which is useful to constrain the peak of the FIR SED and therefore the IR luminosity and dust temperature. Thus, we first perform a FIR SED fitting of AzTEC-1 assuming a combination of a single dust temperature, modified blackbody (MBB) and mid-infrared (MIR) power-law components \citep{2012MNRAS.425.3094C}, and apply the best-fit model of AzTEC-1 as a template to AzTEC-4 and AzTEC-8. To obtain the photometry of 470\,$\mu$m continuum of AzTEC-1, we reduced the Band 9 data (\#\,2018.1.00081.S) and apply a $2''$-aperture. We derived the total flux density of $f_{\rm 470\mu m}=25.01\pm3.16$\,mJy, which is slightly higher than the value reported in \cite{2019ApJ...876....1T}.

We follow the prescription of the single MBB and MIR power-law model described in \cite{2012MNRAS.425.3094C};
\begin{equation}
S(\lambda) = N_{\rm bb}\frac{1-e^{-(\lambda_{0}/\lambda)^{\beta}}}{e^{hc/\lambda kT}-1}\left(\frac{c}{\lambda}\right)^{3} + N_{\rm pl}\left(\frac{\lambda}{\lambda_{c}}\right)^{\alpha}e^{-(\lambda/\lambda_{c})^{2}}\,,
\end{equation}
where $\alpha$, $\beta$, $T$, $\lambda_{0}$, $\lambda_{c}$ are MIR power-law slope, dust emissivity index, dust temperature, the reference wavelength, and MIR power-law turnover wavelength, respectively. The strengths of the MBB and MIR power-law components are related by $N_{\rm BB}$ and $N_{\rm pl}$ through
\begin{equation}
N_{\rm pl}\equiv N_{\rm BB}\frac{1-e^{(\lambda_{0}/\lambda{c})^{\beta}}}{e^{hc/\lambda_{c} kT}-1}\left(\frac{c}{\lambda_{c}}\right)^{3}.
\end{equation}
We adopt the reference wavelength of $\lambda_{0}=200$\,$\mu$m ($\nu_{0}=1500$\,GHz; \citealp{2011ApJ...732L..35C} and the parameters used to constrain the MIR power-law component ($\alpha, \lambda_{c}$) based on the SED fitting of nearby ULIRGs from the GOALS Survey (\citealp{2012MNRAS.425.3094C}; \citealp{2012ApJS..203....9U}). We fix the emissivity index as $\beta= 1.9$. This is a median value of 70 SMGs measured by \cite{2021ApJ...919...30D}, which is in broad agreement with \cite{2023ApJ...951...48M}.
Therefore, the normalization factor $N_{\rm BB}$ and the dust temperature $T$ are the two free parameters during the fitting. We used the {\tt scipy/curve\_fit} function \citep{2020NatMe..17..261V} for the fitting and applied a Monte Carlo method to derive the 16th-84th percentile range as an uncertainty. Having modeled the FIR SED of AzTEC-1, we apply this as a template to AzTEC-4 and AzTEC-8, by normalizing the best-fit SED shape by 860\,$\mu$m flux density after correcting for the effect of redshift. The uncertainties of the IR luminosity are estimated by propagating the uncertainty of flux density ratios. Then, we calculated the dust-obscured SFR ($\rm SFR_{IR}$) following \cite{1998ARA&A..36..189K}:

\begin{equation}
{\rm SFR}_{\rm IR}\,[M_{\odot}{\rm /yr]} = 1.09\times 10^{-10} L_{\rm IR}\,[L_{\odot}]
\end{equation}
after applying the correction factor of 0.63 attributed to the difference in the adopted IMF \citep{2014ARA&A..52..415M}.

\begin{figure*}[t!]
\center
\includegraphics[width=0.9\linewidth]{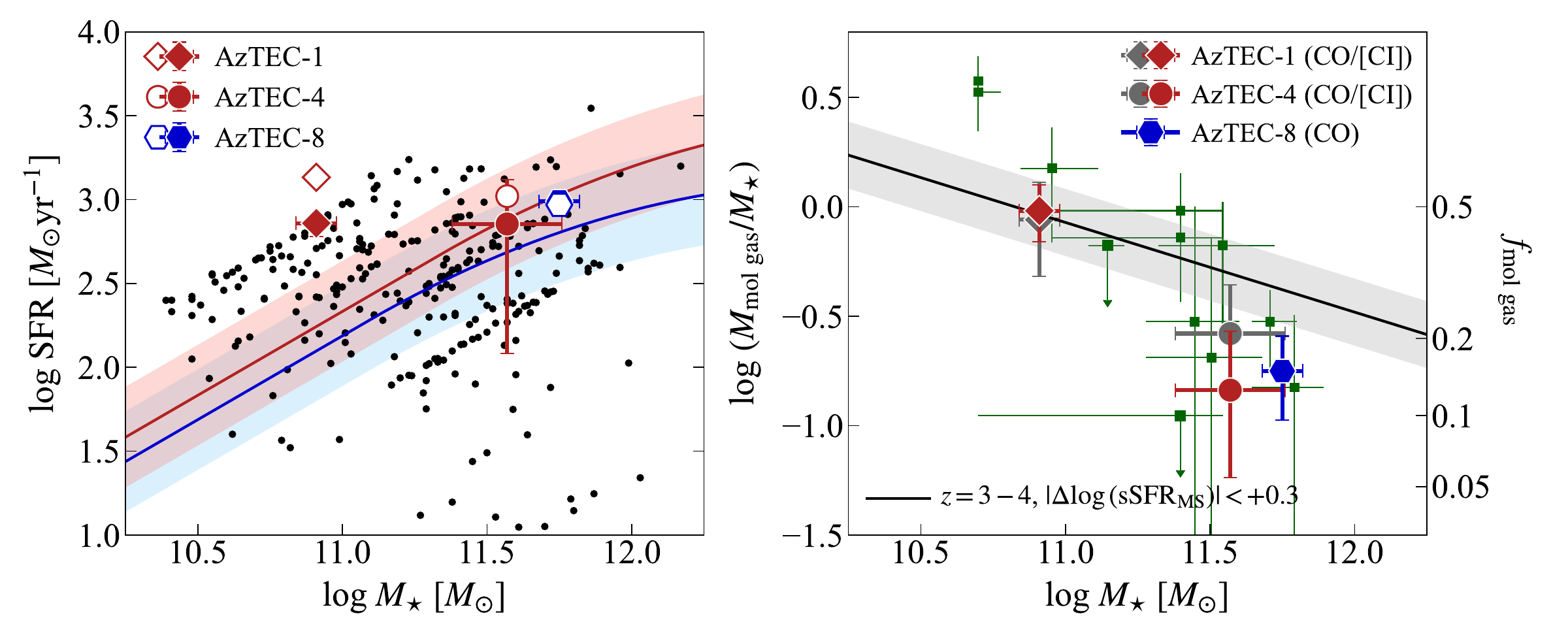}
\caption{{(Left) The stellar mass ($M_{\star}$)-SFR plane of submillimeter-detected galaxies. The filled and open markers correspond to SFRs based on SED fitting and IR luminosity, respectively. The black markers are submillimeter-detected galaxies analyzed in \cite{2025ApJ...979..229M}. The solid lines are the SFMS at $z=3$ (blue) and $z=4$ (red) reported in \cite{2015A&A...575A..74S} with $\pm0.3$\,dex offsets highlighted with the shaded regions. We applied a factor of 1.64 division to the SFMS to correct the difference of IMFs \citep{2014ARA&A..52..415M}. (Right) The molecular gas-to-stellar mass ratio ($M_{\rm mol \ gas}/M_{\star}$) as a function of stellar mass. The gas fraction $f_{\rm mol \ gas}$ is also indicated on the right axis. The solid line and shaded region show a scaling relation for galaxies at $z=3-4$ with $\pm0.3$\,dex offsets from the SFMS \citep{2020ARA&A..58..157T}. For AzTEC-1 and AzTEC-4, both [C\,{\sc i}]-based (red markers) and CO-based (grey markers) measurements are shown. The green markers are SMGs at $3<z<5$ studied in \cite{2023ApJ...945..128F}.}}
\label{fig:Figure7}
\end{figure*}

We list the IR luminosities and dust-obscured SFRs of the three SMGs in Table\,\ref{tab:Table3}. The SED-based SFRs and dust-obscured SFRs are in reasonable agreement except AzTEC-1.  Figure\,\ref{fig:Figure7} shows the stellar mass-SFR plane of three SMGs, compared with other submillimeter-detected galaxies studied in \cite{2025ApJ...979..229M} and star-forming main sequence (SFMS) at $z=3$ and $z=4$ \citep{2015A&A...575A..74S}. Regardless of how the SFRs are derived, AzTEC-1 is located above ($\sim0.7-1.0$\,dex) the SFMS at $z=4$ at fixed stellar mass, suggesting that AzTEC-1 is an extreme starburst galaxy, whereas AzTEC-4 and AzTEC-8 are slightly above but within $0.3$\,dex of the coeval SFMS. We note that the stellar masses reported in \cite{2025ApJ...979..229M} are normalized to match the average $H$-band mass-to-light ratio, which is $\sim$0.2 dex lower than that derived by \cite{2015ApJ...806..110D} and \cite{2020MNRAS.494.3828D}. This discrepancy can be attributed to differences in SED modeling assumptions (e.g., star formation history, dust attenuation properties) and the use of different SED fitting codes. Therefore, care must be taken when comparing the SFMS and other physical properties.

\subsubsection{Molecular gas and dynamical masses}
\label{subsubsec:Section3.1.3}

We estimate the molecular gas and dynamical masses through the different lines detected in the ALMA Band 3 data. Spatially-resolved CO $J=4-3$ line emission in AzTEC-1 is studied in \cite{2018Natur.560..613T} and the detection of CO $J=1-0$ line obtained from the Very Large Array (VLA) observation is further reported in \cite{2023ApJ...945..128F}. Here, we focus on the analysis of ALMA Band 3 data targeting AzTEC-4 and AzTEC-8.

In Figure\,\ref{fig:Figure5} (panels a-1 and b-1), we show the full spectra of AzTEC-4 and AzTEC-8 at 150\,km/s velocity resolution. Three emission lines (CO $J=4-3/5-4$ and \ci$^{3}P_{1}-^{3}P_{0}$) are detected in AzTEC-4, confirming the spectroscopic redshift of AzTEC-4 as $z=4.1979\pm0.0003$. The double-peak profile seen in CO $J=4-3$ and $J=5-4$ lines suggests that AzTEC-4 is a rotating gas disk. Two emission lines (CO $J=3-2/4-3$) are detected in AzTEC-8 ($z=3.097$; \citealp{2022ApJ...929..159C}). The zoom-in view of each emission line in 60\,km/s velocity resolution is also shown in Figure\,\ref{fig:Figure5} (panels a-2, a-3, a-4, b-2, and b-3). To measure the line flux and FWHM, we performed a single Gaussian fitting to each line, using the {\tt scipy/curve\_fit} function. 

Two major uncertainties accompany the derivation of molecular gas mass from rotationally-excited CO lines: excitation ratio and the CO-to-H$_{2}$ conversion factor ($\alpha_{\mathrm{CO}}$). The former is needed to derive the luminosity of ground-state CO emission line (CO\,$J=1-0$), and the latter converts the CO\,$J=1-0$ luminosity into the molecular gas mass, which is primarily composed of molecular hydrogen (H$_{2}$; \citealp{2013ARA&A..51..207B}). For excitation ratios, we adopt the values derived by \cite{2021MNRAS.501.3926B}: $r_{41} = 0.34 \pm 0.04$ for AzTEC-4 and $r_{31} = 0.63 \pm 0.12$ for AzTEC-8, which are broadly consistent with values reported in the literature (e.g., \citealp{2020ApJ...902..109B}; \citealp{2021ApJ...908...95H}; \citealp{2023ApJ...945..128F}).

\begin{deluxetable*}{cccccccccc}[ht!]
\tablecaption{Physical properties \label{tab:Table3}}
\tablehead{
\colhead{Galaxy} & 
\colhead{$f_{860\mu{\rm m},uv}$\tablenotemark{$^{\rm a}$}} & 
\colhead{$f_{860\mu{\rm m},2''}$\tablenotemark{$^{\rm b}$}} & 
\colhead{$M_{\rm\star}$\tablenotemark{$^{\rm c}$}} & 
\colhead{$\rm SFR_{\rm{SED}}$\tablenotemark{$^{\rm c}$}} & 
\colhead{$L_{\rm IR}$} &
\colhead{$\rm SFR_{IR}$} &
\colhead{$M_{\mathrm{mol \,gas, CO}}$} & 
\colhead{$M_{\mathrm{mol \,gas, [CI]}}$} & 
\colhead{$M_{\mathrm{dyn}}$} \\
%\colhead{$f_{\rm gas}$}
%\colhead{$f_{\rm bary}$} \\
\colhead{} & 
\colhead{(mJy)} &
\colhead{(mJy)} &
\colhead{($10^{10}M_{\odot}$)} &
\colhead{($M_{\odot}$/yr)} &
\colhead{($10^{12}L_{\odot}$)} &
\colhead{($M_{\odot}$/yr)} &
\colhead{($10^{10}M_{\odot}$)} & 
\colhead{($10^{10}M_{\odot}$)} & 
\colhead{($10^{10}M_{\odot}$)}
}
%\decimalcolnumbers
\startdata
AzTEC-1 & $17.61\pm0.16$ & $19.48\pm0.80$ & $8.1_{-1.2}^{+1.4}$ & $723\pm120$ & $12.46_{-1.93}^{+2.50}$ & $1358_{-211}^{+273}$ & $7.1\pm3.1$\tablenotemark{$^{\rm d}$} & $7.8\pm1.9$ & $5.68^{+0.64}_{-1.28}$ \\
AzTEC-4 & $12.88\pm0.17$ & $14.52\pm0.71$ & $37.2_{-13.2}^{+20.4}$ & $715\pm594$ & $9.65_{-1.62}^{+2.04}$ & $1048_{-176}^{+221}$ & $9.8\pm3.0$ & $5.4\pm3.0$ & $43.5^{+5.81}_{-11.6}$ \\
AzTEC-8 & $12.20\pm0.09$ & $13.39\pm0.48$ & $56.2_{-8.4}^{+9.8}$ & $977\pm75$ & $8.53_{-1.40}^{+1.78}$ & $931_{-153}^{+194}$ & $10.0\pm3.9$ & -- & $51.4^{+4.59}_{-9.18}$ \\
\enddata
\tablecomments{ {\rm a.} ALMA 860\,$\mu$m continuum flux density measured by modeling visibilities. \\ {\rm b.} 860\,$\mu$m flux density measured by applying $2''$-aperture in the $0\farcs2$-resolution images. \\ {\rm c.} Stellar masses and SFRs derived from the {\tt CIGALE} \citep{2025ApJ...979..229M}. 
\\ {\rm d.} Derived from $I_{\rm{CO(1-0)}}=0.09\pm0.04$ Jy\,km/s measured by \cite{2023ApJ...945..128F}.}
\end{deluxetable*}

It is uncertain which conversion factor is appropriate for our sample, since it may vary for each source depending on gas-phase metallicity and gas surface density. 
For nearby galaxies, it is common to use dichotomous conversion factors, $\alpha_{\rm CO}\sim4$\,$M_{\odot}({\rm K\,km\,s^{-1}\,pc^{-2}})^{-1}$, and $\alpha_{\rm CO}\sim1$\,$M_{\odot}({\rm K\,km\,s^{-1}\,pc^{-2}})^{-1}$ for star-forming spirals and IR-luminous starburst galaxies, respectively \citep{2013ARA&A..51..207B}. \cite{2022MNRAS.517..962D} performed a self-consistent cross-calibration of the conversion factors using three tracers (CO, [C\,{\sc i}], and FIR continuum) for a large sample of high-redshift galaxies. They find that the CO-to-H$_{2}$ and [C\,{\sc i}]-to-H$_{2}$ conversion factors for SMGs as $3.8\pm0.1$ and $16.2\pm0.4$, respectively, which are comparable to the Milky Way value.  On the other hand, \cite{2018ApJ...863...56C}, \cite{2022ApJ...930...35F}, and \cite{2022MNRAS.510.3734D} constrain the CO-to-H$_{2}$ conversion factor of SMGs at $z\sim2-4$ from the dynamical mass derived from gas kinematics, all finding it to be consistent with a value four times smaller than that of the Milky Way. As shown below, using the Milky Way value overestimates the molecular gas mass and even surpasses the dynamical mass. Thus, we adopt the CO-to-H$_{2}$ conversion factor of $\alpha_{\rm{CO}}=1.1$\,$M_{\odot}({\rm K\,km\,s^{-1}\,pc^{-2}})^{-1}$ derived by \cite{2018ApJ...863...56C}. For the [C\,{\sc i}]-to-H$_{2}$ conversion factor, we use the $\alpha_{\rm{[CI]}}=4.4$\,$M_{\odot}({\rm K\,km\,s^{-1}\,pc^{-2}})^{-1}$ ($\alpha_{\rm{[CI]}}/\alpha_{\rm CO}\sim4$ derived by \citealp{2021MNRAS.501.3926B}). Both the CO-based and [C\,{\sc i}]-based molecular gas masses are listed in Table\,\ref{tab:Table3}. {We find that molecular gas mass measurements from CO and [C\,{\sc i}] lines are in reasonable agreement for both AzTEC-1 and AzTEC-4 within the uncertainties. The larger discrepancy in AzTEC-4 is likely due to a combination of uncertainties in the excitation ratio $r_{41}$ and the ratio of conversion factors $\alpha_{\mathrm{[CI]}}/\alpha_{\mathrm{CO}}$. \cite{2025ApJ...987..158F} recently reported the cross-calibrated [C\,{\sc i}]-to-H$_{2}$ conversion factor of $\alpha_{\rm [CI]} = 5.24\pm1.85$\,$M_{\odot}({\rm K\,km\,s^{-1}\,pc^{-2}})^{-1}$ (assuming $\alpha_{\rm CO}=1.0$\,$M_{\odot}({\rm K\,km\,s^{-1}\,pc^{-2}})^{-1}$) for 20 SMGs observed with both [C\,{\sc i}]$^{3}P_{1}-^{3}P_{0}$ and CO\,$J=1-0$ lines. If this conversion factor is applied, the mass discrepancy in AzTEC-4 will be reduced.} To exclude the uncertainty attributed to the excitation ratio, we adopt the [C\,{\sc i}]-based mass for AzTEC-1 and AzTEC-4 hereafter.

For a spherically symmetric rotating system, the dynamical mass can be calculated as $M_{\rm dyn}=v_{\rm rot}^{2}r/G$, where $G$ and $v_{\rm rot}$ are the gravitational constant and rotation velocity, respectively. Substituting the gravitational constant and converting the units yields
\begin{equation}
\label{eq:Equation3}
M_{\rm dyn} = 2.62\times10^{5}v_{\rm rot}r \ [M_{\odot}]
\end{equation}
where $v_{\rm rot}$ and $r$ are in km/s and kpc, respectively. For rotating velocity, we apply $v_{\rm rot}=0.75\,{\rm FWHM}/\sin{i}$, where $i$ is the inclination of the rotating disk and FWHM is measured from the best-fit Gaussian model of emission lines. The factor of 0.75 corresponds to the 20\,\% of the peak flux in a Gaussian profile, which recovers the maximum rotation velocity for quasar-host galaxies at high redshift \citep{2007ApJ...669..821H}. %including 80\,\% of the total flux. 
We use the average $\rm FWHM$ of the CO lines derived from the single Gaussian fitting for AzTEC-4 and AzTEC-8 (Figure\,\ref{fig:Figure5}) and ${\rm FWHM}=305\pm17$\,km/s measured from the CO\,$J=4-3$ line for AzTEC-1 \citep{2018Natur.560..613T}. We estimate the inclination $i$ using the minor-to-major axis ratio ($q$) of either from the resolved CO line emission (AzTEC-1; \citealp{2018Natur.560..613T}) or from the F444W image (AzTEC-4 and AzTEC-8) as $\cos{i}=q$.
%using the minor-to-major axis ratio ($q$) of the F444W image derived in Section\,\ref{subsec:Section3.2} and assuming a ratio of scale height to scale length as $q_{\rm thickness}=0.25$ (Wuyts et al. 2016). The equation for estimating the inclination can be described as
%\begin{equation}
%\label{eq:Equation4}
%\sin{i} = \sqrt{\frac{1-q^{2}}{1-q_{\rm thickness}^{2}}}.
%\end{equation}}

We applied the radius of $r=1.8R_{e}$ to Equation\,\ref{eq:Equation3} which contains 80\,\% of the total flux assuming an exponential disk model. However, the CO effective radius is not available except for AzTEC-1, since the CO lines are not spatially resolved. Thus, for AzTEC-4 and AzTEC-8, we adopt the effective radius measured from the F444W images which both returns S\'{e}rsic index of $n\sim1$ (Section\,\ref{subsec:Section3.2}), as the average size ratio between the CO\,$J=3-2$ line and the 4.4\,$\mu$m continuum emission is close to unity ($R_{e,{\rm CO}}/R_{e,{\rm F444W}}=0.97\pm0.30$) among ten massive DSFGs at $z\sim2$ \citep{2023ApJ...957L..15T}. The dust continuum emission in DSFGs at $z=1-4$ is on average known to be more compact than low-$J$ CO lines by a factor of more than two (\citealp{2022ApJ...933...11I}; \citealp{2025A&A...700A.278R}), therefore is not a suitable tracer of dynamical radius. The resultant dynamical masses are listed in Table\,\ref{tab:Table3}. It has been suggested that assuming a spherically symmetric model may overestimate the dynamical mass by up to 20-30\% compared to a thin disk model or a combination of bulge and disk components (\citealp{2008gady.book.....B}; \citealp{2020MNRAS.497.4051W}), thus we increase the lower uncertainties by a factor of two to take into account this overestimation.

\subsubsection{Baryonic mass and gas mass fractions}
\label{subsubsec:Section3.1.4}

Based on the mass estimations described in Section\,\ref{subsubsec:Section3.1.1} and Section\,\ref{subsubsec:Section3.1.3}, the baryonic mass fraction $f_{\rm bary} = (M_{\star} + M_{\rm mol \ gas})/M_{\rm dyn}$ of three galaxies range $f_{\rm bary}=0.98-2.80$. As dynamical mass reflects the total mass enclosed within the radius $r$ ($1.8R_{e}$ in our calculation), the dynamical mass should be larger than the baryonic mass, i.e., $f_{\rm bary} < 1$. However, only AzTEC-4 ($f_{\rm bary}=0.98$) satisfies this criterion. This unphysical outcome is occasionally reported in the literature and commonly attributed to the large uncertainties in the mass measurements (e.g., \citealp{2020Natur.581..269N}; \citealp{2021ApJ...917...94M}; \citealp{2022MNRAS.510.3734D}).

The stellar mass measurements of AzTEC-4 and AzTEC-8 based on SED fitting have large uncertainties, presumably because their rest-frame UV continuum is not detected. In turbulent gas disks, the contrast in radial pressure between the central core and the outer disk becomes pronounced, leading to a stronger radial pressure gradient. As demonstrated by \cite{2020MNRAS.497.4051W} using simulated galaxies in the FIRE simulations (\citealp{2014MNRAS.445..581H}; \citealp{2018MNRAS.480..800H}), the radial pressure gradient can underestimate the intrinsic rotational velocity and therefore the dynamical mass by up to $40$\,\%.\footnote{{This is because a negative radial pressure gradient induces a force between radii $r=r_{0}$ and $r=r_{0}+\Delta r$, which acts against the gravitational force. For more details, see \cite{2010ApJ...725.2324B} and \cite{2020MNRAS.497.4051W}.}} Therefore, we argue that the unphysically high baryonic mass fractions of the three SMGs can be largely attributed to either the overestimation of the stellar mass or the underestimation of the dynamical mass.

{The right panel of Figure\,\ref{fig:Figure7} shows the molecular gas-to-stellar mass ratio as a function of stellar mass for the three SMGs, as well as for SMGs at $3<z<5$ studied in \cite{2023ApJ...945..128F}.} AzTEC-1 has a gas mass fraction, defined as $f_{\rm mol \ gas}= M_{\rm mol\ gas}/(M_{\star}+M_{\rm mol \ gas})$, of $f_{\rm mol \ gas}=0.47\pm0.22$, which is in agreement with the scaling relation from \citep{2020ARA&A..58..157T}, whereas {AzTEC-4 and AzTEC-8 have $f_{\rm mol \ gas}\lesssim0.2$, which are $\sim0.3-0.5$\,dex below the scaling relation. SMGs with similarly low gas fractions (using the CO\,$J=1-0$ line) have also been reported by \cite{2023ApJ...945..128F}. Nonetheless,} if the stellar masses of AzTEC-4 and AzTEC-8 are overestimated, their gas mass fractions may be consistent with the expected scaling relation.

To summarize, AzTEC-1 is a starburst galaxy $\sim0.7-1.0$\,dex above the SFMS {reported in \cite{2015A&A...575A..74S}}, whereas AzTEC-4 and AzTEC-8 {lie} on the massive end of, or slightly above the SFMS, although their stellar mass measurements are uncertain. The dynamical masses derived from the FWHM of the CO lines are consistent with the baryonic masses within uncertainties for all galaxies except AzTEC-1, implying a different dynamical state for AzTEC-1 compared to the other two galaxies. Spatially-resolved gas kinematics is indispensable for further confirming this finding.

\subsection{Morphological properties}
\label{subsec:Section3.2}

\begin{figure}
\center
\includegraphics[width=0.95\columnwidth]{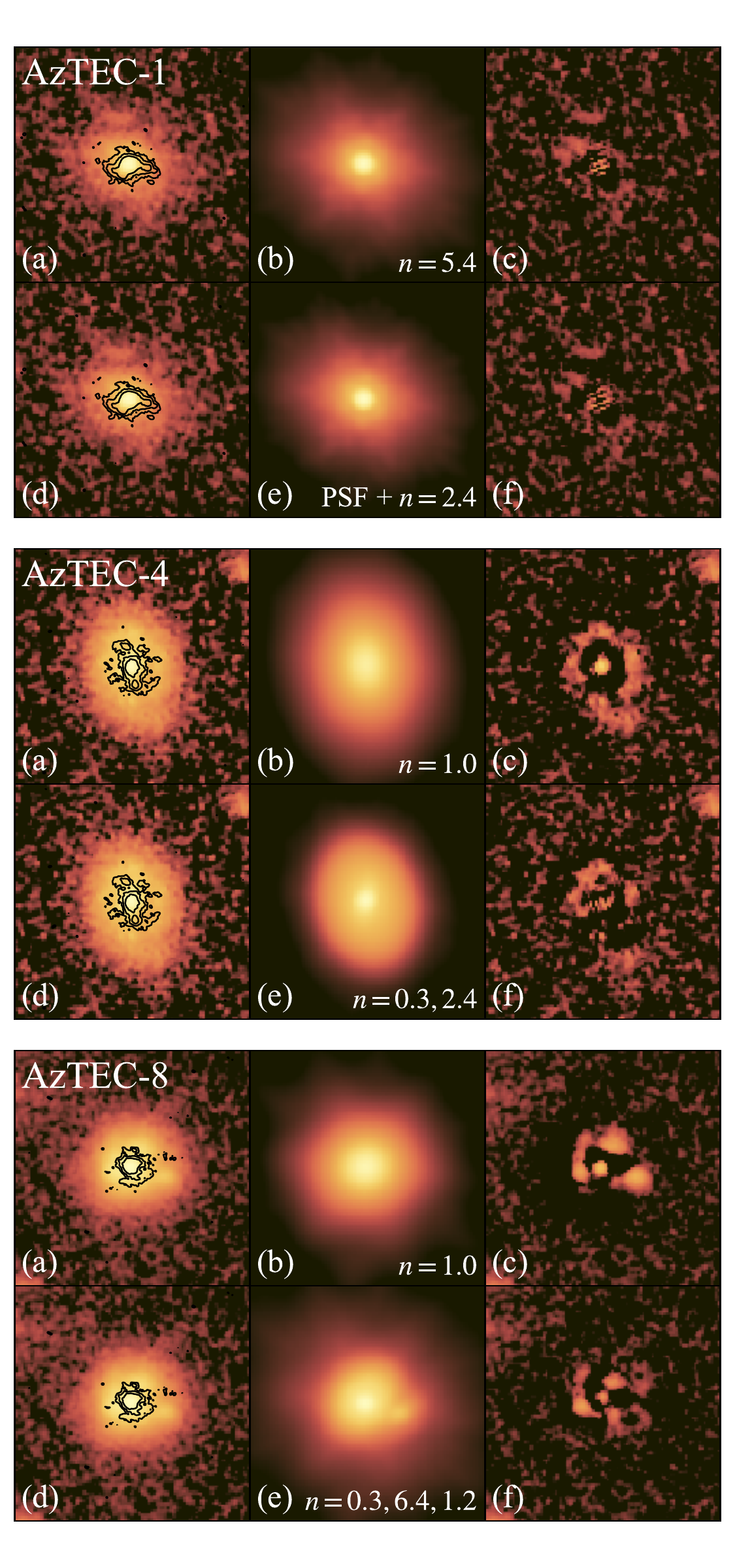}
\caption{Fitting results of {\tt GALFIT} at F444W filter. From left to right, each panel shows the input (panels a and d), fitted model (panels b and d), and residual (panels c and f) maps within the $3''\times3''$ region. The maps are all shown in logarithmic scale. The fitted S\'{e}rsic indices are shown in the lower-right corner of the model image. An off-centered companion galaxy candidate with $n=1.2$ is added in the three components fit of AzTEC-8. Black contour levels on the input image indicate $[3, 6, 12]\,\sigma$ of ALMA 860\,$\mu$m continuum.\label{fig:Figure8}}
\end{figure}

We quantify the morphological properties using {\tt GALFIT}  \citep{2002AJ....124..266P} which performs parametric fitting of objects in two-dimensional images. We applied the {\tt GALFIT} analyses to JWST F444W images assuming that the rest-frame optical emission can be well represented by a S\'{e}rsic profile \citep{1968adga.book.....S}. We run {\tt GALFIT} by feeding a $15''\times15''$ cutout science image, standard deviation  (ERR extension) image produced by the JWST pipeline, and the PSF image. All three SMGs returns the best-fit single S\'{e}rsic model of the F444W images. For AzTEC-1, {\tt GALFIT} returned a compact and concentrated profile with an effective radius of $R_{e,\rm major}=1.4$ kpc and a S\'{e}rsic index of $n=5.4$. The single fit leaves a point-like source at the center of the residual map (Figure\,\ref{fig:Figure8}). As we will discuss later, it is possible that AzTEC-1 is in a transition phase between a dusty starburst and an optically-bright quasar. Therefore, we also performed a fit using a combination of a point source and a single S\'{e}rsic profile, where the former corresponds to the quasar component. This yields a flatter S\'{e}rsic component with $R_{e,\rm major}=1.8$ kpc and $n=2.4$. In contrast to AzTEC-1, {\tt GALFIT} returned profiles close to an extended exponential disk ($R_{e,\rm major}\sim2-3$\,kpc, $n\sim1$) for AzTEC-4 and AzTEC-8. 

We present the morphological modeling results for the F444W images in Figure\,\ref{fig:Figure8}. The first rows (panels a, b, and c) show the results assuming a singel S\'{e}rsic model. As reported in the literature, a single S\'{e}rsic component is occasionally insufficient to model the stellar structure of massive DSFGs at high redshift (e.g., \citealp{2022ApJ...939L...7C}; \citealp{2023ApJ...942L..19C}). The presence of significant residuals in the AzTEC-4 and AzTEC-8 images prompted us to further fit double and triple S\'{e}rsic components for AzTEC-4 and AzTEC-8, respectively. The third S\'{e}rsic component of AzTEC-8 is modeled as a non-concentric component, as indicated by the residual map of the single S\'{e}rsic model. These results are shown in the second row of Figure\,\ref{fig:Figure8} (panels d, e, and f). As expected, these multi-component fit returns smaller reduced chi-square ($\chi^{2}$) values than the single component fit. We provide the summary of the results in Table\,\ref{tab:Table4}. The single-component S\'{e}rsic indices vary between compact `bulge'-like component ($n\sim4$) and extended `disk'-like component ($n\sim1$).

\begin{figure*}[ht!]
\includegraphics[width=\linewidth]{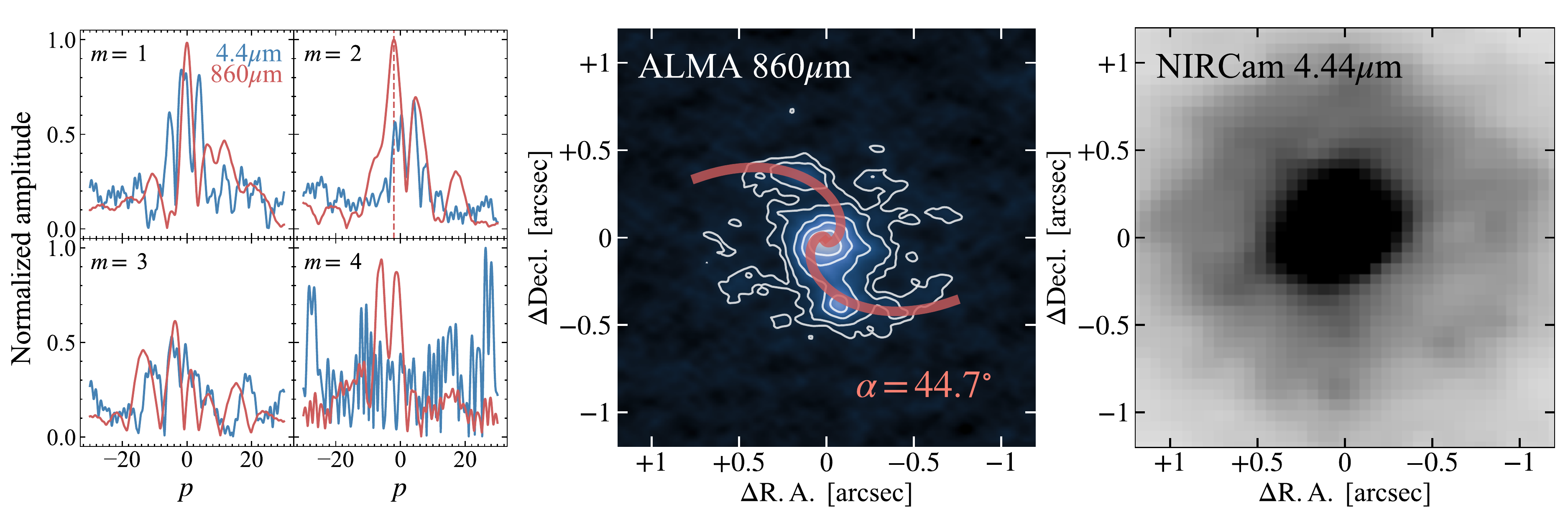}
\caption{AzTEC-4, a two-arm dusty spiral galaxy at $z=4.198$. Left panel shows a normalized amplitude of four ($m=1-4$) Fourier modes of 4.4\,$\mu$m (blue) and 860\,$\mu$m (red) continuum images as a function of a dimensionless parameter $p=-m/\tan\alpha$. The strongest peak is seen in $m=2$ mode for 860\,$\mu$m continuum image, indicating a two-arm spiral structure with a pitch angle of $44.7^{\,\circ}$. Middle and right panels show 860\,$\mu$m and 4.4\,$\mu$m continuum images deprojected to a face-on view. The red spiral overlaid on the 860\,$\mu$m image corresponds to the strongest Fourier component in the $m=2$ mode. The contour levels of [3, 5, 10, 15, 20]\,$\sigma$ are shown.
\label{fig:Figure9}}
\end{figure*}

\begin{deluxetable*}{cccccc}
\tablecaption{Morphological properties} \label{tab:Table4}
\tablewidth{0pt}
\tablehead{
\colhead{Galaxy} & 
\colhead{$\lambda_{\rm{obs}}$} & 
\colhead{S\'{e}rsic components} & 
\colhead{$R_{e,{\rm major}}$} & 
\colhead{$n$} & 
\colhead{$q$} \\
\colhead{} & \colhead{($\mu$m)} & \colhead{} & \colhead{(kpc)} & \colhead{} & \colhead{}
}
%\decimalcolnumbers
\startdata
\multirow{4}{*}{AzTEC-1} & \multirow{2}{*}{4.44} & single & $1.38\pm0.02$ & $5.43\pm0.10$ & $0.74\pm0.01$ \\
\cline{3-6}
& & PSF $+$ single & $1.79\pm0.02$ & $2.41\pm0.07$ & $0.72\pm0.01$ \\
\cline{2-6}
 & 860 & -- & $1.19\pm0.02$ & 1 (fixed) & $0.64\pm0.01$ \\
\cline{2-6}
 & 2060 & -- & $0.75\pm0.03$ & 1 (fixed) & $0.66\pm0.04$ \\
\hline
\multirow{4}{*}{AzTEC-4} & \multirow{3}{*}{4.44} & single & $2.68\pm0.01$ & $1.02\pm0.01$ & $0.74\pm0.00$ \\
\cline{3-6}
&  & \multirow{2}{*}{double} & $3.14\pm0.01$ & $0.27\pm0.00$ & $0.74\pm0.00$ \\
& & & $0.87\pm0.01$ & $2.42\pm0.21$ & $0.52\pm0.01$ \\
\cline{2-6}
& 860 & -- & $1.49\pm0.03$ & 1 (fixed) & $0.61\pm0.02$ \\
\hline
\multirow{6}{*}{AzTEC-8} & \multirow{4}{*}{4.44} & single & $1.91\pm0.00$ & $0.96\pm0.00$ & $0.87\pm0.00$ \\
\cline{3-6}
&  & \multirow{3}{*}{triple} & $2.14\pm0.01$ & $0.26\pm0.00$ & $0.80\pm0.00$ \\
& & &$1.99\pm0.06$ & $6.42\pm0.25$ & $0.76\pm0.01$ \\
& & & $0.93\pm0.01$ & $1.24\pm0.05$ & $0.48\pm0.01$  \\
\cline{2-6}
& 860 & -- & $0.83\pm0.01$ & 1 (fixed) & $0.92\pm0.02$ \\
\cline{2-6}
& 1270 & -- & $0.61\pm0.06$ & 1 (fixed) & $0.92\pm0.11$
\enddata 
\end{deluxetable*}

\subsubsection{Detection of the two-arm spiral of dust at z = 4.2}
\label{subsec:Section3.2.1}

Motivated by the intriguing morphologies of three SMGs shown in Figure\,\ref{fig:Figure1}, we perform a two-dimensional Fourier analysis to investigate whether the three SMGs host a spiral structure. We follow the method presented by \cite{1992A&AS...93..469P}, which was originally proposed by \cite{1977ApJ...212..637K} and has been mostly applied to nearby galaxies to characterize the multi-arm spiral structure (e.g., \citealp{2012ApJS..199...33D}; \citealp{2019A&A...631A..94D}; \citealp{2020ApJ...900..150Y}). \cite{2021Sci...372.1201T} applied this method to spatially-resolved \cii line emission in an intensely star-forming optically-bright quasar, BRI\,1335-0417 at $z=4.41$ observed by ALMA and detected an evidence of two-arm spiral structure. A radial profile of an $m$-mode logarithmic spiral can be described as $r=r_{0}\exp[-(m/p)\theta]$ in the polar coordinates ($r$, $\theta$), where $p$ is the dimensionless parameter which characterizes the pitch angle ($\alpha$) of a spiral through $p=-m/\tan(\alpha)$. Fourier analysis decomposes the image into Fourier amplitude $A$ determined by each $m$ and $p$:
\begin{equation}
A(m,p) = \frac{1}{D}\sum_{j} d_{j} \exp[-i(pu_{j}+m\theta_{j})] , 
\end{equation}
where subscript $j$ signifies each pixel, $u_{j}\equiv \ln r_{j}$, $d_{j}$ is a weight of each pixel corresponding to the flux density, and $D=\sum_{j} d_{j}$. The relative strength of the Fourier amplitudes provides information about which mode and pitch angle of the spiral structure is dominant in the images. In this paper, we will compare the Fourier amplitudes normalized by the strongest peak within the $m=1-4$ modes.

Galaxies inclined toward the line of sight tend to resemble a two-arm spiral structure as they approach an orientation of edge-on view, thus we first need to correct for the inclination to obtain a face-on view. We estimate the inclination ($i$) of the galaxies based on the minor-to-major axis ratio ($q$) of the F444W image measured by {\tt GALFIT} by assuming that the stellar disk is thin and circular, i.e., $q = \cos i$. We then deproject the galaxies to a face-on view by adjusting their aspect ratios along the minor axis while preserving the pixel scale and the beam size, using the {\tt AffineTransform} function implemented in {\tt scikit-image} \citep{2014PeerJ...2..453V}. We subsequently calculate the Fourier amplitudes for modes $m = 1$--$4$, with $p$ sampled from $-30$ to $30$.

We find that the 860\,$\mu$m continuum of AzTEC-4 shows the strongest Fourier amplitude in $(m, p)=(2, -2.0)$, corresponding to a two-arm spiral with a pitch angle of $\alpha=44.7^{\circ}$ (Figure\,\ref{fig:Figure9}). The peak of the $m=1$ mode with $\alpha\sim-90^{\circ}$ in 860\,$\mu$m continuum is also fairly strong, likely reflecting the asymmetry between the northern and southern arms of AzTEC-4. On the other hand, we do not find strong signature of $m=2$ or $m=3$ modes in the 4.4\,$\mu$m continuum emission. For AzTEC-1 and AzTEC-8, the strongest Fourier amplitude are found in either $m=1$ or $m=4$ in 4.4\,$\mu$m and 860\,$\mu$m continua.

%Based on the strongest peak at the $m=2$ mode in 860\,$\mu$m continuum emission of AzTEC-4, we report a detection of the two-arm dusty spiral at $z=4.198$. 
The strong $m=2$ signal obtained from the 860\,$\mu$m continuum emission of AzTEC-4 suggests a two-arm dusty spiral at $z=4.198$. 
Several studies reported dusty spirals and bars at high redshift, identified either through the morphology of NIRCam images (\citealp{2022ApJ...938L..24F}; \citealp{2023ApJ...942L...1W}; \citealp{2023ApJ...958L..26H}; \citealp{2024A&A...690A.285P}; \citealp{2025MNRAS.536.3090K}), high-resolution ALMA observations (\citealp{2021Sci...372.1201T}; \citealp{2025A&A...693L..17S}; \citealp{2025MNRAS.540L..78D}), or a combination of both (\citealp{2025PASJ...77..432U}; \citealp{2025ApJ...978..165H}; \citealp{2025Natur.641..861H}). AzTEC-4 represents the first case of high redshift galaxies where we see a two-arm spiral in dust, but not in the stellar distribution. Ignoring differences in components, the loose pitch angle of AzTEC-4 may be consistent with the redshift evolution of pitch angles of spiral arms observed by HST and JWST (\citealp{2023A&A...680L..14R}; \citealp{2025PASA...42...29C}).  

Studies based on numerical simulations suggest a spontaneous formation of a two-arm spiral structure either in a galaxy pair during a major merger (\citealp{1996ApJ...464..641M}; \citealp{2004ApJ...616..199I}), in a tidal encounter caused by a minor merger \citep{2017MNRAS.468.4189P}, or in a disk galaxy in isolation by the classical density wave theory, local instabilities, and swing amplification (e.g., \citealp{2014PASA...31...35D}; \citealp{2022ARA&A..60...73S}). In particular, \cite{2022ApJ...934...52B} present a $z = 2$ galaxy with a two-armed spiral structure in the gaseous component, driven by a central bar, from a zoom-in cosmological simulation using {\tt GIZMO} \citep{2017arXiv171201294H}. The stellar distribution does not exhibit this feature, which is reminiscent of AzTEC-4. \cite{2022ApJ...934...52B} highlight that all of their simulated bars are triggered by interactions.

While we are unable to determine the spatial extent of molecular gas disk with the current data, \cite{2014ApJS..214....1U} argue that molecular gas disks which are more extended than the stellar bulge may be produced after a major merger, possibly leading to a formation of a gravitationally unstable gas-rich disk \citep{2009ApJ...703..785D}. Assuming that the molecular gas disk traced by CO\,$J=3-2$ line has a similar spatial extent as the stellar distribution \citep{2023ApJ...957L..15T}, AzTEC-4 would have a central gas surface density of $\log\Sigma_{\rm mol \ gas}=\log(M_{\rm mol \  gas}/2\pi R_{e}^{2})= 2.52-3.21$\,$M_{\odot}/{\rm kpc}^{2}$, which is close to or above the threshold for an unstable gas disk with a Toomre $Q$ parameter larger than unity \citep{2024A&A...691A.299G}.

According to major merger simulations, \cite{2008ApJ...688..972C} argue that SMGs with disk-like morphologies at rest-frame optical wavelengths could have experienced a major merger in the past. However, the simulations only reproduced a peak 850\,$\mu$m flux density as bright as $\sim4-5$\,mJy. \cite{2010MNRAS.401.1613N} present the major merger simulations reaching $\sim15$\,mJy at comparable mass, but the simulated SMGs show a disturbed optical morphology even 600\,Myr after the final coalescence. Since the optical morphology of AzTEC-4 appears relatively undisturbed and no clear companion galaxies based on the photometric redshifts from the COSMOS2025 catalog were found within 30\,kpc radius from AzTEC-4, we favor the scenario where its intense star formation is primarily driven by gravitationally instability of a gas-rich disk (\citealp{2023ApJ...942L..19C}; \citealp{2024A&A...691A.299G}; \citealp{2024A&A...690A.285P}; \citealp{2024ApJ...969...27B}). 

Finally, hydrodynamic $N$-body simulations conducted by \cite{2024ApJ...968...86B} show that galaxies with low gas fraction can preserve stellar bars and spiral arms for more than 1 Gyr. The simulated gas bars  are only intermittent, which arise from young stars on bar-like orbits. This may explain why we do not see a two-arm spiral feature in optical wavelengths, as dust obscuration plays an effective role in attenuating UV light from young stellar populations. High-resolution gas kinematics with a comparable resolution to 860\,$\mu$m continuum will be a key to verify the formation mechanism of the two-arm dust spiral.

\section{Discussion and summary} 
\label{sec:Section4}

\begin{figure*}[t!]
\center
\includegraphics[width=\linewidth]{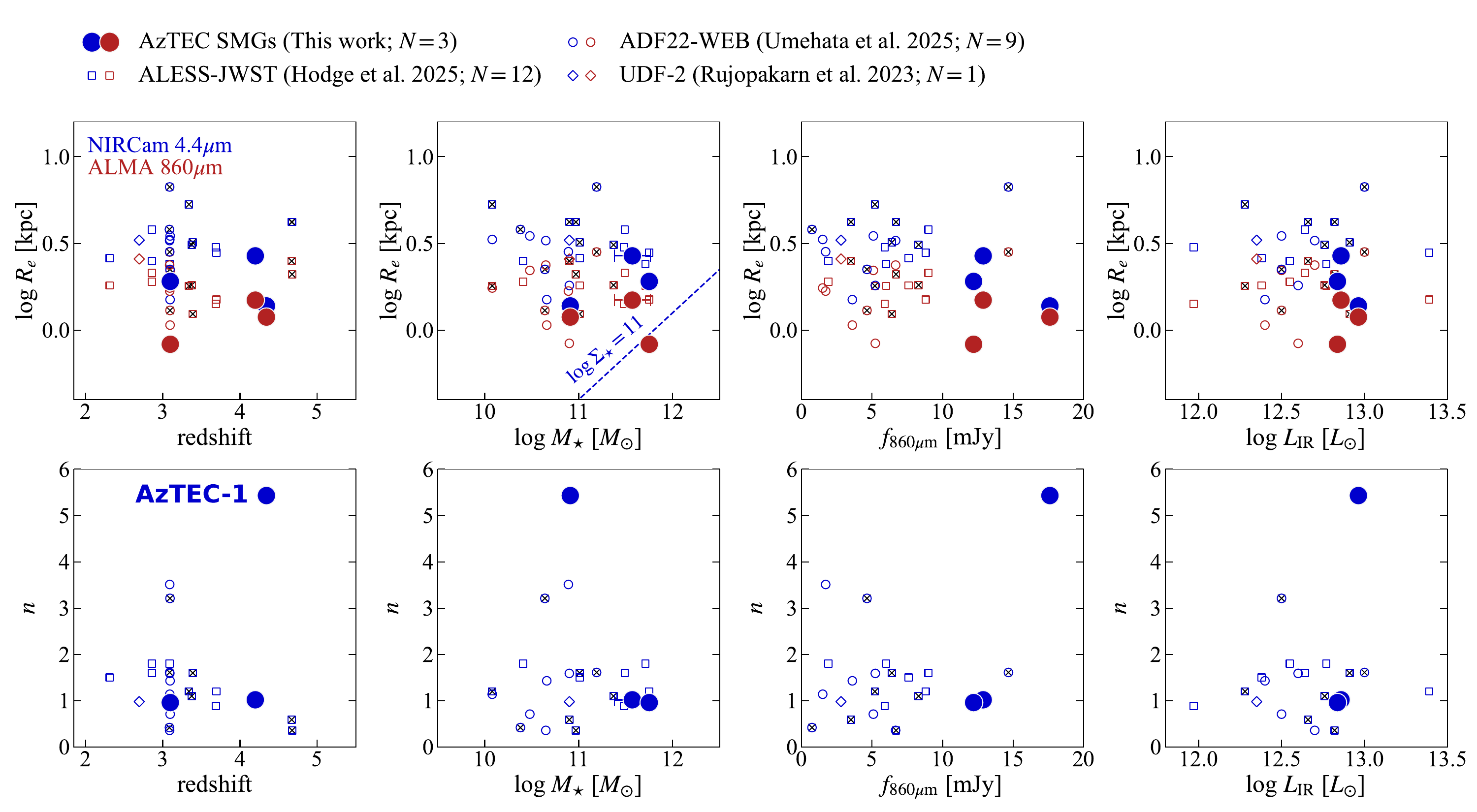}
\caption{Physical properties of the three SMGs. Top and bottom panels show the effective radius and S\'{e}rsic index based on a single S\'{e}rsic model as a function of four different properties (redshift, stellar mass, 860$\mu$m flux density, and IR luminosity). Red and blue markers indicate the measurements from the ALMA 860\,$\mu$m and JWST\,$4.44$\,$\mu$m images, respectively. Other datapoints are the compilation of SMGs at $z>2$ with spatially-resolved ALMA 860\,$\mu$m and JWST\,$4.44$\,$\mu$m images (\citealp{2023ApJ...948L...8R}; \citealp{2025ApJ...978..165H}; \citealp{2025PASJ...77..432U}). The third column from the left shows that the three AzTEC sources are among the brightest SMGs, with $f_{\rm 860\mu m} > 12$\,mJy. Crosses indicate galaxies that are visually identified to be mergers. The measurements used in this figure is retrieved from \cite{2015ApJ...806..110D}, \cite{2017ApJ...835...98U}, \cite{2021ApJ...919...30D}, and \cite{2025A&A...699A.324H}. For the IR luminosity of ALESS SMGs, we use the updated measurements of \cite{2021ApJ...919...30D} if available. \label{fig:Figure10}}
\end{figure*}

We have explored the global and morphological properties of three luminous SMGs, which were originally identified by JCMT/AzTEC and followed up in high angular resolution by ALMA and JWST/NIRCam. Despite the similarities in submillimeter fluxes and SFRs, we find diverse morphological properties among three SMGs. AzTEC-1 has a concentrated structure ($n=5.4$) in the optical wavelengths and a clumpy morphology in the FIR wavelengths which shows at least three FIR clumps with no clear optical counterparts. The optical morphology of AzTEC-4 and AzTEC-8 can be characterized by a disky profile, suggesting a fundamental difference in the nature of these galaxies.

Figure\,\ref{fig:Figure10} compares the effective radius and \sersic index as a function of four different properties: redshift, stellar mass ($M_{\star}$), 860\,$\mu$m flux density ($f_{860\mu{\rm m}}$), and IR luminosity ($L_{\rm IR}$). In addition to the three SMGs, we have compiled 22 SMGs and DSFGs at $z>2$ which have been studied in high-resolution images in both rest-frame optical and FIR wavelengths (\citealp{2023ApJ...948L...8R};  \citealp{2025ApJ...978..165H}; \citealp{2025arXiv250201868U}). Among them, nine SMGs were visually classified as mergers (\citealp{2025ApJ...978..165H}; \citealp{2025arXiv250201868U}). The SMGs studied in this paper represent the most luminous sources ($f_{\rm 860\mu {\rm m}}>12$\,mJy) among the compiled sample, with AzTEC-4 and AzTEC-8 being among the most massive galaxies ($\log (M_{\star}/M_{\odot})>11.5 $). AzTEC-8 has the highest stellar mass surface density of $\log(\Sigma_{\star}/M_{\odot}{\rm kpc^{-2}})=\log(M_{\star}/2\pi R_{e,{\rm F444W}}^2)=10.35\pm0.07$, although not as high as the maximum stellar mass surface density of $\log(\Sigma_{\star}/M_{\odot}{\rm kpc^{-2}})=11$, which is likely the limit regulated by feedback from massive stars \citep{2010MNRAS.401L..19H}. Nonetheless, as AzTEC-4 and AzTEC-8 both share similar parameter space in Figure\,\ref{fig:Figure10} (except for 860\,$\mu$m flux density) as other SMGs in the literature, and show no morphological evidence of ongoing merger activity, it is likely that they have undergone secular growth through continuous gas accretion.

From the bottom panels of Figure\,\ref{fig:Figure10}, it is clear that AzTEC-1, despite its high redshift, has an extraordinarily high \sersic index among the compiled SMG sample. AzTEC-1 also has the smallest effective radius of F444W. What makes AzTEC-1 unique among the SMG sample in terms of these properties? Another intriguing fact of AzTEC-1 is that the spatially-resolved \cii kinematics reveals two non-corotating gas clumps \citep{2020ApJ...889..141T} and a kinematic asymmetry along the major axis \citep{2023MNRAS.521.1045R}. \cite{2020ApJ...889..141T} argue that these non-corotating gas clumps were formed ex-situ and were the cause of a violent disk instability, while the possibility that AzTEC-1 is a spatially unresolved merger cannot be ruled out \citep{2023MNRAS.521.1045R}. The smaller dynamical mass compared to the stellar and molecular gas masses (Section\,\ref{subsubsec:Section3.1.3}) suggests that AzTEC-1 is not a dynamically virialized system. In contrast, as we have seen in Figure\,\ref{fig:Figure1}, there is no clear evidence for on-going merging activity in the optical images taken by JWST.

We suggest that AzTEC-1 is a major merger remnant of gas-rich galaxies. The highly concentrated and compact stellar morphology with a substantial amount of molecular gas disk ($f_{\rm mol \ gas}=0.47\pm0.22$) support this scenario, indicating that the formation of the stellar core was triggered by the efficient loss of angular momentum of the cold gas. The star-forming disk traced by FIR continuum emission can be formed in a short timescale after a major merger between two gas-rich disk galaxies (\citealp{2005ApJ...622L...9S}; see also \citealp{2019MNRAS.485.1320M}). A remnant of gas-rich major mergers simulated in \cite{2005ApJ...622L...9S} reserve a significant gas fraction and is able to form a rotationally supported gas disk and a multiplex stellar structure reminiscent of late-type galaxies. The high SFR can sustain even after the coalescence phase of a major merger if the cold gas accretion from the circumgalactic medium continues, which may be observed through \cii and FIR clumps \citep{2014ApJ...780...57B}. Based on the fractional 860\,$\mu$m fluxes compared to the total flux of the three FIR clumps around the stellar core of AzTEC-1, each clump is expected to have a mass of approximately $\sim 3\times10^{9}\,M_{\odot}$. These properties are consistent with the in situ clumps, which will survive for hundreds of Myr, found in theoretical simulations (\citealp{2014ApJ...780...57B}; \citealp{2014MNRAS.443.3675M}). 
Lastly, we note that the formation of a compact stellar bulge with an extended gas disk is compatible with the scenario that cold gas inflow triggers the starburst \citep{2009ApJ...703..785D}, but if this is the common mechanism, then we likely have observed similar structure in other SMGs more frequently. Therefore, we argue that the rarity of major merger events compared to minor mergers at high redshift (e.g., \citealp{2021MNRAS.501.3215O}) makes AzTEC-1 unique among the SMG population studied in high resolution to date.

The compact morphology and unobscrued nature in the rest-frame UV of AzTEC-1, unlike other SMGs, are reminiscent of optically-bright quasars. The stellar mass surface density of $\log(\Sigma_{\star}/M_{\odot}{\rm kpc^{-2}})=9.83 \pm 0.07$ is comparable or even higher than quasar-host galaxies \citep{2022ApJ...939L..28D}. In concordance with the above discussion that AzTEC-1 is a major merger remnant, we suggest that AzTEC-1 is in a transition phase between a SMG and a dust-obscured quasar \citep{2010MNRAS.407.1701N}. Similarly, an SMG, ASXDF1100.057.1 ($z_{\rm phot}=1.9^{+0.11}_{-0.04}$), which was classified as an AGN-dominated galaxy based on MIR color-color selection, has been reported to exhibit a point-like morphology in the rest-frame UV and extended FIR emission \citep{2017ApJ...849L..36I}. Thus, although a larger sample is necessary to draw definitive conclusions, these morphological features may serve as useful classifiers of SMGs with different origins.

The emerging picture after the launch of JWST is that majority of the SMGs and DSFGs have disky structure in rest-frame optical wavelengths, with some fraction of them hosting sub-structures, such as stellar clumps, bars, and spirals (\citealp{2022ApJ...939L...7C}; \citealp{2023ApJ...942L..19C}; \citealp{2023ApJ...958...36S}; \citealp{2024A&A...691A.299G}; \citealp{2024A&A...690A.285P}; \citealp{2024ApJ...969...27B}; \citealp{2025ApJ...978..165H}; \citealp{2025arXiv250201868U}; \citealp{2025Natur.641..861H}). This is in stark contrast to the major merger-driven starburst scenario supported in the pre-ALMA era both in observations (\citealp{2003ApJ...596L...5C}; \citealp{2010ApJ...724..233E}; \citealp{2010MNRAS.406..230R}) and in theoretical simulations (e.g., \citealp{2011ApJ...735L..34G}). In this paper, we have presented that multiple physical origins exist for triggering starburst in luminous SMGs at high redshift, by analyzing the combined images of $\sim500$\,pc resolution images of rest-frame optical and FIR continuum emission in three SMGs beyond redshift three. We found that AzTEC-1 ($z=4.34$) is a post major merger, analogous to nearby ultra-luminous infrared galaxies. The compact rest-frame UV/optical morphology suggests that it may represent the initiating phase of a dust-obscured quasar, which is followed by an optically bright phase after the surrounding dust has been blown out, consistent with the `major merger' paradigm that posits mergers trigger quasar activity and eventually lead to the formation of massive quiescent galaxies. AzTEC-4 ($z=4.20$) is likely a gravitationally unstable gas disk, which hosts a two-arm dusty spiral structure but not in the stellar emission. The nature of AzTEC-8 ($z=3.10$) remains unclear; however, it is considered to follow either of the following scenarios: a starburst triggered by a gravitationally unstable gas disk, an ongoing minor merger, or a combination of both. A larger SMG sample with resolved FIR and optical images down to sub-kpc resolutions is needed to determine the prevalence of each mechanism.

\begin{acknowledgments}
We thank the anonymous referee for constructive comments.
This paper makes use of the following ALMA data:ADS/JAO.ALMA \#2015.1.01345.S, \#2016.1.00012.S, \#2017.A.00034.S, \#2017.1.00127.S, \#2017.1.00487.S, \#2018.1.00081.S, \#2018.1.01136.S, and \#2019.1.01600.S. ALMA is a partnership of ESO (representing its member states), NSF (USA) and NINS (Japan), together with NRC (Canada), MOST and ASIAA (Taiwan), and KASI (Republic of Korea), in cooperation with the Republic of Chile. The Joint ALMA Observatory is operated by ESO, AUI/NRAO and NAOJ. We thank the ALMA staff and in particular the EA-ARC staff for their support. This work is also based in part on observations made with the NASA/ESA/CSA James Webb Space Telescope. R.I. is supported by Grants-in-Aid for Japan Society for the Promotion of Science (JSPS) Fellows (KAKENHI Grant Number 23KJ1006). D.I. is supported by Grants-in-Aid for Japan Society for the Promotion of Science (JSPS) Fellows (KAKENHI Grant Number 23K20870). M.F. acknowledges the funding from the European Union's Horizon 2020 research and innovation programme under the Marie Sk{\l}odowska-Curie grant agreement No.101148925. Y.T. is supported by Grants-in-Aid for Japan Society for the Promotion of Science (JSPS) Fellows (KAKENHI Grant Number 22H04939, 23K20035, 24H00004). T.M. is supported by Grants-in-Aid for Japan Society for the Promotion of Science (JSPS) Fellows (KAKENHI Grant Number 25K17441). M.L. acknowledges the support from the European Union’s Horizon 2020 research and innovation program under the Marie Sk{\l}odowska-Curie grant agreement No.101107795. 
{Some of the data presented in this article were obtained from the Mikulski Archive for Space Telescopes (MAST) at the Space Telescope Science Institute. The specific observations analyzed can be accessed via \dataset[doi:10.17909/a1gq-c588]{https://doi.org/10.17909/a1gq-c588}.} These observations are associated with JWST Cycle 1 GO program \#1727. Data analyses were in part carried out on the Multi-wavelength Data Analysis System operated by the Astronomy Data Center (ADC), National Astronomical Observatory of Japan.
\end{acknowledgments}
	

%\facilities{HST(STIS), Swift(XRT and UVOT), AAVSO, CTIO:1.3m, CTIO:1.5m,CXO}


\software{ALMA Interforometric Pipeline \citep{2023PASP..135g4501H}; {\tt astropy} \citep{2022ApJ...935..167A}; {\tt CASA} \citep{2022PASP..134k4501C}; {\tt GALFIT} \citep{2002AJ....124..266P}; %{\tt photoutils} \citep{2023zndo...1035865B}; 
{\tt PSFEx} \citep{2011ASPC..442..435B}; {\tt pypher} \citep{boucaud2016}; Scientific colour maps \citep{2021zndo...1243862C}; {\tt scikit-image} \citep{2014PeerJ...2..453V}; {\tt UVMULTIFIT} \citep{2014A&A...563A.136M}}



%\begin{appendix}
%\restartappendixnumbering
%\section{Morphological analysis}
%\label{sec:AppendixA}

%\section{Description of AzTEC-8 field}
%\label{sec:AppendixB}

%\end{appendix}

\begin{comment}
\begin{figure}
\includegraphics[width=0.95\columnwidth]{figures/AzTEC8_field.pdf}
\caption{AzTEC-8 field.}
\end{figure}

\end{comment}

\bibliography{aztec_ikeda}{}
\bibliographystyle{aasjournalv7}



\end{document}